\documentclass[aip,showpacs,superscriptaddress]{revtex4}
\usepackage{amsmath,amssymb,amsfonts}
\usepackage{bbold}
\usepackage{color}
\usepackage{url}
\usepackage{graphicx}
\usepackage{tikz}
\usepackage{epstopdf}

\usepackage{hyperref}

\newcommand{\eq}[1]{Eq.~(\ref{#1})}
\newcommand{\noeq}[1]{~(\ref{#1})}

\newcommand{\Refs}[1]{Refs.~[\onlinecite{#1}]}
\newcommand{\leb}[1]{\text{d}[#1]}
\newcommand{\dd}[1]{\text{d}#1}
\newcommand{\Det}{\text{det}\,}

\newcommand{\Sdet}{\text{sdet}\,}
\newcommand{\Str}{\text{str}\,}
\newcommand{\tr}{\text{tr}\,}
\newcommand{\id}{\mathbb{1}}
\newcommand{\IR}{\mathbb{R}}

\newcommand{\diag}{\text{diag}\,}

\newcommand{\str}{\text{str}\,}
\newcommand{\sdet}{\text{sdet}\,}

\begin{document}

\title{Asymptotic Coincidence of the Statistics for Degenerate and
  Non--Degenerate Correlated Real Wishart Ensembles}

\author{Tim Wirtz}\email[]{tim.wirtz@iais.fraunhofer.de}
\affiliation{Fakult\"at f\"ur Physik, Universit\"at Duisburg--Essen,
  Duisburg, Germany} 
\affiliation{Fraunhofer Institut f\"ur
  Intelligente Analyse-- und Informationssysteme, Sankt Augustin,
  Germany} 
\author{Mario Kieburg}\email[]{kieburg@physik.uni-bielefeld.de}
\affiliation{Fakult\"at f\"ur Physik, Universit\"at Duisburg--Essen,
  Duisburg, Germany}\affiliation{Fakult\"at f\"ur Physik,
  Universit\"at Bielefeld, Bielefeld, Germany} 
\author{Thomas Guhr} \email[]{thomas.guhr@uni-due.de} 
\affiliation{Fakult\"at f\"ur Physik,
  Universit\"at Duisburg--Essen, Duisburg, Germany}


\newcommand{\corr}[1]{{\color{red}#1}}
\newcommand{\tim}[1]{{\color{blue}#1}}

\date{\today}

\begin{abstract}
  The correlated Wishart model provides the standard benchmark when
  analyzing time series of any kind. Unfortunately, the real case,
  which is the most relevant one in applications, poses serious
  challenges for analytical calculations. Often these challenges are
  due to square root singularities which cannot be handled using
  common random matrix techniques. We present a new way to tackle this
  issue. Using supersymmetry, we carry out an anlaytical study which
  we support by numerical simulations. For large but finite matrix
  dimensions, we show that statistical properties of the fully
  correlated real Wishart model generically approach those of a
  correlated real Wishart model with doubled matrix dimensions and
  doubly degenerate empirical eigenvalues.  This holds for the local
  and global spectral statistics.  With Monte Carlo simulations we
  show that this is even approximately true for small matrix
  dimensions. We explicitly investigate the $k$--point correlation
  function as well as the distribution of the largest eigenvalue for
  which we find a surprisingly compact formula in the doubly
  degenerate case. Moreover we show that on the local scale the
  $k$--point correlation function exhibits the sine and the Airy
  kernel in the bulk and at the soft edges, respectively. We also
  address the positions and the fluctuations of the possible outliers
  in the data.
\end{abstract}

\pacs{05.45.Tp, 02.50.-r, 02.20.-a}

\maketitle 

\section{Introduction}

Random matrix theory was first introduced in biostatistics by
Wishart~\cite{Wishart} and later on also by Wigner in the context of
Hamiltonian systems~\cite{Wigner1951,Wigner1967}. It has extraordinary
power to model and study generic features in a variety of systems, see
Ref.~\cite{Guhr1998}. It only employs basis invariance and global
symmetries of the matrices resulting in the orthogonal, unitary and
symplectic ensembles~\cite{dyson}. Wishart's ideas opened a new
direction in time series analysis and statistical
inference~\cite{muirhead,Anderson2003,chatfield,Johnstone,BS}. The
Wishart model is widley used, including applications in fields such as
medicine~\cite{Seba}, biophysics\cite{pianka2011evolutionary},
chemistry\cite{Feinberg1979},
finance~\cite{LalouxCizeauBouchaudPotters,ple02}, wireless
communication~\cite{TulinoVerdu}, to mention just a few. The Wishart
model shares the unique advantage of all random matrix approaches:
Most of its predictions are accessible in experiments or observations
and can therefore directly be tested. Although the random matrix
theory setup is straightforward, calculations are often difficult. The
real case which is the most relevant one for applications is
particularly cumbersome.

We thus focus on the case of $p\times n$ rectangular matrices $W$ with
real entries $W_{j\nu}\in \IR$ for $j=1,\dots,p$ and $\nu=1,\dots,n$.
The $p$ rows of $W$ may be viewed as model time series of length $n$. We
assume a Gaussian distribution~\cite{muirhead,Anderson2003},
\begin{align}
P(W|C) \sim \exp\left(-\frac{ n}{2}\tr~WW^T C^{-1}\right) \ ,
\label{WishartDist}
\end{align}
where the $p\times p$ matrix $C$ is the empirical correlation matrix
specific for the data under consideration. This matrix is input of the
model and requires to be real symmetric with positive eigenvalues
$\Lambda_i$, $i=1,\dots,p$. In particular we have $C=V\Lambda V^T$
with $V\in \text{O}(p)$ and
$\Lambda=\diag(\Lambda_1,\dots,\Lambda_p)$. The positive definite
$p\times p$ matrix $WW^T$ is the model correlation matrix and
due to our choice of $P(W|C)$, it is on average $\langle
WW^T\rangle=C$.

In applications of the real Wishart model, correlated or not, square
roots of characteristic polynomials and therefore branch cuts
arise. For instance, gap probabilities related to the smallest and
largest eigenvalue were found to possess a representation as averaged
products of determinants in the denominator to half integer
power~\cite{WirtzGuhrKieburgEPL2015}. Other examples are the
eigenvalue density in the ordinary and doubly correlated Wishart
model~\cite{VP,RecheretalPRL,Recheretal,WaltnerWirtzGuhr2014}, the
distribution of the smallest
eigenvalue~\cite{WirtzGuhrII,WirtzGuhrI,Akemannetal2013,Akemannetal2014}
as well as universality considerations in scattering
theory~\cite{FyodorovKhoruzhenkoNock2013,FyodorovNock2014}. Those
square roots are serious obstacles in analytical calculations and a
solution is urgently called for. To the best of our knowledge a
comprehensive analytical strategy for averages over a product of
characteristic polynomials to half integer power does not exist. For
certain special cases some solutions are
known~\cite{Akemannetal2013,Akemannetal2014,FyodorovNock2014}.

The analytical calculations drastically simplify in the case that the
empirical correlation matrix becomes doubly degenerate, because the
square roots are not present anymore. Although this case is
empirically rarely justified, our results provide very good
approximations for the case without such degeneracies. Our main goal
is a general approach to eigenvalue statistics in the real correlated
Wishart model, which to some extent outmanoeuvers the square roots of
characteristic polynomials such that standard random matrix techniques
apply. Based on analytical calculations using
supersymmetry~\cite{SUSYZirnbauer,GuhrSUSY} and on numerical
simulations we verify that most of the statistical properties in the
bulk, at the edges and for the outliers of an arbitrary, correlated
real Wishart ensemble do not depend on the degree of the degeneracy of
the empirical correlation matrix.  In particular, the spectral
observables of a $p\times n$ random matrix $W$ correlated with $C$
coincide with those of an $lp\times ln$ random matrix correlated with
$C\otimes\id_l$ where $l\in\mathbb{N}$ is the degree of degeneracy and
$\id_l$ is the $l$-dimensional identity matrix. This statement becomes
exact for
\begin{align}
0 \ < \ \frac{p}{n} \ = \ \gamma^2 \ < \ 1 \ \ll \ n,p \ \to \ \infty
\end{align}
under very moderate assumptions on the empirical correlation matrix
$C$. The eigenvalue density of correlated Wishart ensembles with non-degenerate spectrum was already studied by many others in \cite{MarcenkoPastur,SilversteinChoi,silversteinBai,Shinzato}. We will regain their results and additionally we derive results about the local spectral statistics.

As a by--product we also derive the sine and the Airy kernel for real
matrices in the bulk and at the soft edges, respectively, for the
fully correlated case. Importantly, we properly account for all
Efetov--Wegner boundary contributions~\cite{PS1979,Wegner,efetov}
which often pose substantial difficulties in supersymmetry
calculations. To this end we apply Rothstein's theory~\cite{Roth} and
identify the results with those for the Gaussian Orthogonal Ensemble
(GOE). We include outliers and discuss their positions and
fluctuations, provided they are well--separated from all other
eigenvalues.

We show that most of the spectral observables are independent of the
degree of degeneracy, and we thus claim that the distributions of the
largest eigenvalue for the correlated real Wishart ensembles with the
empirical correlation matrices $C$ and $C\otimes\id_2$ are
approximately equal in the limit of large matrix dimensions $p$ and
$n$. We derive a representation of the cumulative density function in
terms of a $p\times p$ Pfaffian ($p$ even) for the $2p\times 2n$
Wishart ensemble with $C\otimes\id_2$. For this purpose we start from
an earlier result~\cite{WirtzGuhrKieburgEPL2015} and employ
skew--orthogonal polynomials~\cite{AKP10}.

Although the results are derived in an asymptotic limit, we find
surprisingly good agreement with numerical simulations already for
rather small matrix dimensions. This allows a quantitative as well as
a qualitative spectral analysis in the Wishart model without doublely
degenerate empirical eigenvalues if $p/n\sim\mathcal{O}(1)$ and $n,p$
large.

Our study is structured as follows. In section~\ref{superRep}, we
summarize the basics of the $k$-point function and present the
corresponding supermatrix model. We also discuss the conditions on $C$
to ensure that the limit $n,p\to\infty$ with $p/n\in[0,1]$ is
well--defined. The saddle point approximation of the supermatrix model
is performed in section~\ref{sec:bulk} in which we also derive a
simple general relation between the macroscopic level density
(marginal density) and the saddle point solution. Furthermore, we study the
bulk and the edges of the spectrum and derive the sine kernel on the
local scale. In section~\ref{sec:saddlepoint-edge} we investigate
possible outliers and the local statistics of the soft edges and
derive the Airy kernel. We also manage to express the cumulative
distribution of the largest eigenvalue in terms of a Pfaffian
determinant. For illustrating purpose and to confirm our claims, we perform
numerical simulations in section~\ref{numerics}. We conclude in
section~\ref{conclusion}. A brief sketch of Rothstein's
theory~\cite{Roth} is relegated to appendix~\ref{app:Rothstein}.

\section{Supersymmetric Representation of the $k$--Point Correlation
  Functions}
\label{superRep}

The $k$--point correlation function $R_k(x;\xi)$ with $k\leq p$
measures the eigenvalue fluctuations of the model correlation matrix
$WW^T$. We use two sets of variables $x=\diag(x_1,\dots,x_k)$ and
$\xi=\diag(\xi_1,\dots,\xi_k)$ for later separation of the global and
the local scales, respectively.  To study the local scale, we unfold
the spectrum with the level density which depends on the empirical
eigenvalues $\Lambda=\diag(\Lambda_1,\ldots,\Lambda_p)$. Importantly,
on the original and on the unfolded scale, all $k$--point functions may
depend non--trivially on these empirical eigenvalues. One of the main
results to be derived below is the emergence of the universal
statistical features of the \textit{uncorrelated} Wishart ensemble
after unfolding and under modest conditions on $\Lambda$. Furthermore,
an arbitrary degeneracy of degree $l\in\mathbb{N}$ of the empirical eigenvalues
$\Lambda\to\Lambda\otimes\id_l$ does not change the statistics. Even
the global level density $R_1(x)$ remains the same for large matrix
dimensions $n,p\to\infty$. In sections~\ref{supmat}
and~\ref{testlimits} we set up the supermatrix model and test the
asymptotics, respectively.

\subsection{Setting up the supermatrix model}
\label{supmat}

To be as general as possible, we consider an ensemble of Wishart
matrices $W$ of size $lp\times ln$ drawn from the normal
distribution\noeq{WishartDist}, where the eigenvalues of $C$ are
$l$-fold degenerate, \textit{i.e.} the empirical eigenvalues are
$\Lambda\otimes\id_l$. For this ensemble we analyze its $k$-point
correlation function which is expressed as the derivative of
a generating function,
\begin{align}
\label{Rk}
 R_k(x;\xi ) =\left. \frac{1}{(4 \pi \imath)^k (lp)^k }\sum_{L\in\{\pm 1\}^k}\prod_{i=1}^k L_i \partial_{j_i}Z_{k,k}^{(p,n)}(\kappa)\right|_{\substack{\varepsilon\rightarrow0\\j=0}}~,
\end{align}
where $\kappa_{b,1} = x_b + j_b + \xi_b/(lp) + \imath L_b
\varepsilon$, $\kappa_{b,2} = x_b - j_b + \xi_b/(lp) + \imath L_b
\varepsilon$ and $L_b=\pm1$ for $b=1,\dots,k$. The generating function
also depends on source variables $j=\diag(j_1,\dots,j_k)$. The scaling
of the variables $\xi_a$ with $lp$ anticipates the local scale for
spectral  fluctuations inside the macroscopic bulk in which the
unscaled variables $x_a$ are assumed to lie. The latter variables
$x_a$ may also be degenerate, {\it i.e.}~$x_a=x_b$ for some
$a,b=1,\ldots,p$, as long as the eigenvalues $x_a+\xi_a/(lp)$ are
pairwise different. The scaling of $\xi_a$ has to be adjusted when one
or more of the variables $x_a$ are at an edge of the spectrum.
The generating function reads
\begin{align}
 \label{GenerFunc}
 Z_{k,k}^{(p,n)}(\kappa) = \int\leb{W} P(W| C)~ \prod_{b=1}^{k}\frac{\Det\left(WW^T - \kappa_{b,2}\id_{lp}\right)}{\Det\left(WW^T - \kappa_{b,1}\id_{lp}\right)}~,
\end{align}
with $\leb{\cdot}$ being the flat measure, \textit{i.e.}~the product
of all independent differentials. The matrix $\id_{lp}$ is the
$lp$ dimensional identity matrix.

To conveniently study the asymptotics for large $n,p$ with
$p/n=\gamma^2$ fixed, we employ the supersymmetry method, see
Refs.~\cite{efetov,Verbaarschotetsal,guhr1991,GuhrSUSY,SUSYZirnbauer}. A
more mathematical introduction into superanalysis can be found in
Ref.~\cite{Ber}. Using the results in
\Refs{Recheretal,WirtzGuhrI,KaymakKieburgGuhr2014}, we map the
generating function\noeq{GenerFunc} to superspace,
\begin{align}
\begin{split}
\label{superZ}
 Z_{k,k}^{(p,n)}(\kappa) =& K_{nl,k} \sdet^{(n-p)l/2}\tilde\kappa \int\leb{\sigma}~ \sdet^{(nl-1)/2}\sigma \exp\left(\imath\frac{nl}{2}\str\tilde\kappa \sigma\right)\sdet^{-l/2}\left(\id_p\otimes\id_{2k|2k} +\imath\Lambda\otimes\sigma\right),
\end{split} 
\end{align}
where $\tilde \kappa =
\diag(\kappa_{1,1},\ldots,\kappa_{k,1},\kappa_{1,2},\ldots,\kappa_{k,2})\otimes\id_2$
is viewed as a $(2k|2k)\times(2k|2k)$ diagonal supermatrix. The second superdeterminant  corresponds to the Gaussian factor which would occur naturally for the Gaussian orthogonal ensemble, cf. Eq.~\eqref{GOE-k-point}. However we consider the correlated real Wishart ensemble which yields a different weight factor in superspace.
The factor
of $2$ in the dimensions occurs because we study the \textit{real}
correlated Wishart ensemble. The $(2k|2k)\times (2k|2k)$ supermatrix
$\tilde L\sigma$ has a positive definite symmetric matrix in the
boson--boson block $\sigma_{\rm BB}$ while the fermion--fermion block
$\sigma_{\rm FF}$ belongs to the circular symplectic
ensemble~\cite{Zirnbauer1996,Sommers2007,LittlemanSommersZirnbauer,KaymakKieburgGuhr2014}.
The boson--fermion block $\sigma_{\rm
  BF}=\{\eta_{ab},\eta_{ab}^*\}_{a=1,\ldots,2k; b=1,\ldots,k}$
consists of $2k$ real independent Grassmann variables and the
fermion-boson block is $\sigma_{\rm FB}=-\sigma_{\rm BF}^\dagger$ with
the dagger denoting the ordinary adjoint.  Here we have employed the
supermatrix $\tilde L =
\diag(L_{1},\ldots,L_{k})\otimes\id_{1|1}\otimes\id_2$ encoding the
signs of the imaginary increment $\varepsilon$. The normalization
constant
\begin{flalign}
\begin{split}
\label{superK}
 K_{nl,k}^{-1} = & \int\leb{\sigma} \sdet^{(nl-1)/2}\sigma \exp\left(-\frac{nl}{2}\str \tilde L\sigma\right),
\end{split} 
\end{flalign}
is determined by the condition that $Z_{k,k}^{(p,n)}\rightarrow1$ for
$\varepsilon\rightarrow\infty$. By construction, we also have
$Z_{k,k}^{(p,n)}(\kappa)|_{j=0}=1$ for vanishing source variables.  To
show the non--trivial equality of the integral~\eqref{superK} for the
normalization constant and the integral~\eqref{superZ} for $j=0$, one
needs Cauchy--like integral
theorems~\cite{PS1979,efetov,Con,ConGro,KKG08} first derived by
Wegner~\cite{Wegner} for arbitrary supermatrix sizes.
The measure $\leb{\sigma}$ is the product of all differentials of the
independent variables. The integration over Grassmann variables
are normalized as
\begin{equation}
\int d\eta=0,\quad\int \eta d\eta=1 \ ,
\end{equation}
which differs from another convention by a factor of $\sqrt{2\pi}$.
With this choice the constant $K_{nl,k}$ becomes in the large $n$ limit
\begin{equation}\label{asymp-const}
K_{\infty,k}=\lim_{n\to\infty}K_{nl,k}= 4^{-k}(2\pi^2)^{-k^2}
\end{equation}
because the integrand can be expanded around $\sigma_0=\tilde{L}$
yielding a Gaussian integral.

In the supermatrix representation~\eqref{superZ} we differentiate with
respect to the source variables $j_a$ and set them to zero. Then we
perform a $1/p$ expansion by means of a saddle point approximation. We
expand around the saddle point matrix $\sigma_0$ according to
$\sigma=\sigma_0+\delta\sigma/\sqrt{p}$ where the scaling $\sqrt{p}$
of the massive modes $\delta\sigma$ is dictated by the fact that all
variables $x_a$ are in the bulk of the spectrum. After keeping only
the leading order term we find
\begin{align}
 \label{eq:dd:kpointfunction}
 \begin{split}
 R_k(x,\xi) = &K_{nl,k}\lim_{\varepsilon\rightarrow0} 
 \int\leb{\sigma_0,\delta\sigma} \sum_{L_1,\ldots,L_k=\pm 1}\exp\left(-\frac{nl}{2}\mathcal{L}\left(\sigma_0+\frac{\delta\sigma}{\sqrt{p}}\right)\right)\\
 &\times\prod_{j=1}^k\left(\frac{L_j }{8\pi \gamma^2}\str\left(\sigma_0+\frac{\delta\sigma}{\sqrt{p}}\right)\left[\begin{array}{cc} e^k_{jj}&0\\0&- e^k_{jj}\end{array}\right]\otimes\id_{2}+\frac{\gamma^{-2}-1}{2\pi\imath}\frac{L_j}{x_j+\imath L_j\varepsilon}\right)+\mathcal{O}\left(\frac{1}{p}\right)
 \end{split}
\end{align}
with $\gamma^2=p/n$. Here $e^k_{ab}$ is a $k\times k$ matrix with
zeros everywhere and unity in the $(a,b)$ entry. For the time being,
neither the saddle point manifold of $\sigma_0$, referred to as
Goldstone modes, nor the support of the massive modes $\delta\sigma$
are precisely specified.  The second term $1/(x_j+\imath
L_j\varepsilon)$ in the above product is reminiscent of the
superdeterminant in front of the integral~\eqref{superZ}. It generates
Dirac $\delta$ functions $\delta(x_j)$ which have the following origin:
To derive the expression~\eqref{superZ} we used $WW^T$ instead of
$W^TW$. Their spectra only differ in the number of the generic zero
eigenvalues which is equal to $n-p=(\gamma^{-2}-1)p$ for $W^TW$ and
zero for $WW^T$.  We return to these terms in
subsection~\ref{sec:Level}.  Keeping with the common terminology,
we refer to the function
\begin{align}
\label{approxLagrangian}
 \begin{split}
\mathcal L (\sigma) =&  \frac{\gamma^2}{p}\sum_{i=1}^{p} \Str\ln\left(\id_{2k|2k}+\imath \Lambda_i \sigma\right)-\imath\Str\left(\tilde x+\imath\varepsilon\tilde L+\frac{\tilde\xi}{pl}\right) \sigma - \left(1-\frac{\gamma^2}{pl}\right) \str\ln \sigma.
 \end{split}
\end{align}
in the above expression as ``Lagrangian''.

\subsection{Testing the limit of large matrix dimensions}
\label{testlimits}

We now show that the limit $p,n\to\infty$ with $0<\gamma^2=p/n\leq1$
fixed is well--defined, because $|\Sdet^{-1}\left(1+\imath \Lambda_i
  \sigma\right)|$ is bounded.  For the numerical part $\hat\sigma$ of $\sigma$ we have
\begin{equation}\label{SUSY-ev}
\left|\Sdet^{-1}\left(\id_{2k|2k}+\imath \frac{\Lambda_i}{2\Lambda_{\max}} \hat\sigma\right)\right|=\frac{\prod_{j=1}^k|1+\imath\Lambda_i e^{\imath\varphi_j}/(2\Lambda_{\max})|^2}{\prod_{j=1}^{2k}|1+\imath L_j\Lambda_i e^{\theta_j}/(2\Lambda_{\max})|} \ ,
\end{equation}
where
$e^\theta=\tilde{L}\diag(e^{\theta_1},\ldots,e^{\theta_{2k}})/(2\Lambda_{\max})$
are the eigenvalues of the boson--boson block $\sigma_{\rm BB}$ of
$\sigma$ and
$e^{\imath\varphi}=\diag(e^{\imath\varphi_1},\ldots,e^{\imath\varphi_{k}})\otimes\id_2/(2\Lambda_{\max})$
are the eigenvalues of the fermion--fermion block $\sigma_{\rm
  FF}$. Here we rescaled $\sigma\to \sigma/(2\Lambda_{\max})$ with
$\Lambda_{\max}$ being the largest of the empirical eigenvalues
$\Lambda$. The expression~\eqref{SUSY-ev} is bounded from below and
above according to
\begin{equation}
0<\frac{(1-\Lambda_i/(2\Lambda_{\max}))^{2k}}{\prod_{j=1}^{2k} (1+ e^{2\theta_j}\Lambda_i/(2\Lambda_{\max}))}\leq\left|\Sdet^{-1}\left(\id_{2k|2k}+\imath \frac{\Lambda_i}{2\Lambda_{\max}} \hat\sigma\right)\right|\leq\left(1+\frac{\Lambda_i}{2\Lambda_{\max}}\right)^{2k}<\infty.
\end{equation}
These bounds are integrable due to the terms $\exp[-nl\varepsilon
e^{\theta_j}/2]$ and $e^{(nl-1)\theta_j/2}$ and due to $k\leq p\leq n$
in the integrand. This estimate only holds for the part of
$\prod_{i=1}^p|\Sdet^{-1}\left(\id_{2k|2k}+\imath \Lambda_i
  \sigma\right)|$ without the Grassmann variables. An expansion in the
Grassmann variables yields a finite polynomial in powers of the
matrices
\begin{equation}
\frac{1}{p}\sum_{i=1}^{p}\frac{1}{\left(2\Lambda_{\max}/\Lambda_i\id_{2k}+\imath\sigma_{\rm BB}\right)^{\otimes m}}\otimes\frac{1}{\left(2\Lambda_{\max}/\Lambda_i\id_{2k}+\imath\sigma_{\rm FF}\right)^{\otimes m}}
\end{equation}
which are contracted in the generating function~\eqref{superZ}, the
details do not matter. The tensor product multiplies the space
corresponding to the $2k\times 2k$ boson--boson block with the one
corresponding to the $2k\times 2k$ fermion--fermion block. The
exponent $m=0,\ldots,2k^2$ is taken in a tensor sense, too. The
modulus of the spectrum of these matrices are bounded from above by
$2^{-2m}$ independent of $\Lambda$ and $\hat\sigma$. Therefore the
limit $p,n\to\infty$ with $0<\gamma^2=p/n\leq1$ fixed is well--defined
if we assume that
\begin{equation}\label{assumption}
 \lim_{p\to\infty}\left|\frac{1}{p}\sum_{i=1}^{p}{\rm ln}(1+s \Lambda_i)\right|<\infty
\end{equation}
remains finite for any $s>-1/\Lambda_{\rm max}$ and in the case that
$\Lambda_{\rm max}/\Lambda_i$ also remains finite. This is realized
when the smallest eigenvalue is of the same order as the largest eigenvalue
$\Lambda_{\rm max}$.
 
If $\Lambda$ contains a finite number $p_{\rm out}$ of outliers of
larger order as the ones in the bulk, we may still resort to the
discussion above. We split the product of superdeterminants in two
parts,
\begin{equation}\label{split-outlier}
 \prod_{i=1}^p\sdet^{-l/2}\left(\id_{2k|2k} +\imath\Lambda_i\sigma\right)= \prod_{i=1}^{p-p_{\rm out}}\sdet^{-l/2}\left(\id_{2k|2k} +\imath\Lambda_i\sigma\right)\prod_{i=p-p_{\rm out}+1}^{p}\sdet^{-l/2}\left(\id_{2k|2k} +\imath\Lambda_i\sigma\right).
\end{equation}
Only the first product enters the saddle point equation to be given in
the sequel while the second one may be considered as a
$p$--independent perturbation of the integrand. The second product
cannot contribute to the saddle point analysis since the number of
outliers $p_{\rm out}$ is assumed to be fixed. The physical
interpretation is that outliers which are macroscopically separated
from or may even lie on a scale larger than that of the bulk do not
influence the statistics in the bulk. We study the outliers in more
detail in subsection~\ref{sec:outlier}.

Another remark is in order, clarifying how the existence of a limiting
distribution $\rho(\lambda)$ for the empirical eigenvalues $\Lambda$
affects the above discussion. Such a distribution exists if
\begin{equation}\label{lim-dis.a}
\lim_{p\to\infty} \frac{1}{p}\sum_{i=1}^{p}f(\Lambda_i)=\int\limits_0^\infty f(\lambda)\rho(\lambda){\rm d}\lambda,
\end{equation}
whenever the test function $f$ is integrable with respect to $\rho$
and $f(\Lambda_i)<\infty$. The sum in the
Lagrangian~\eqref{approxLagrangian} is then bounded from above and below by the average of the
integral of the supermatrix resolvent,
\begin{equation}\label{lim-dis.b}
\lim_{p\to\infty} \frac{1}{p}\sum_{i=1}^{p}  \str\ln\left(\id_{2k|2k}+\imath \Lambda_i \sigma\right)=\int\limits_0^\infty \str\ln\left(\id_{2k|2k}+\imath \lambda \sigma\right)\rho(\lambda){\rm d}\lambda.
\end{equation}
Outliers appear as Dirac $\delta$ functions in $\rho$.  Although
Eqs.~\eqref{lim-dis.a} and \eqref{lim-dis.b} are only valid if a
limiting distribution for the empirical eigenvalues $\Lambda$ exists,
we want to find an expression which still provides a good
approximation at finite matrix dimensions $p$ and $n$.

\section{Bulk Statistics}\label{sec:bulk}

We analyze the bulk statistics in three steps. First we discuss the
saddle point approximation for a general $k$-point correlation
function in section~\ref{saddlepoint}.  In section~\ref{sec:Level} an
explicit and very simple relation between the macroscopic level
density and the saddle point solution is presented.  In
section~\ref{sec:k-point} we show that inside the bulk the whole
spectral statistics on the local scale agrees with the sine kernel of
real matrices.  This is true and exact for all $k$--point correlation functions
including the cumbersome Efetov--Wegner boundary terms, see
Refs.~\cite{Guhr06,Kieburg2011}.

\subsection{Saddle point approximation in the bulk}\label{saddlepoint}

We now show that, assuming the condition~\eqref{assumption}, the
$k$--point correlation function is independent of the degree $l$ of
degeneracy in leading order of a $1/p$ expansion. To this end, we carry out a saddle
point approximation of the integral\noeq{eq:dd:kpointfunction} by
expanding the Lagrangian~\eqref{approxLagrangian} up to the order
$1/p$,
\begin{align}
\label{approxLagrangian.b}
 \begin{split}
\mathcal L \left(\sigma_0+\frac{\delta\sigma}{\sqrt{p}}\right) =&  \frac{\gamma^2}{p}\sum_{i=1}^{p} \Str\ln\left(\id_{2k|2k}+\imath \Lambda_i \sigma_0\right)-\imath\Str\left(\tilde x+\imath\varepsilon\tilde L\right) \sigma_0 - \left(1-\frac{\gamma^2}{pl}\right) \str\ln \sigma_0\\
&+\frac{\imath}{\sqrt{p}}\Str\left[ \frac{\gamma^2}{p}\sum_{i=1}^{p} \frac{\Lambda_i}{\id_{2k|2k}+\imath \Lambda_i \sigma_0}-\tilde x-\imath\varepsilon\tilde L +\imath \sigma_0^{-1}\right]\delta\sigma\\
&+\frac{1}{2p}\Str\left[\frac{\gamma^2}{p}\sum_{i=1}^{p} \left(\frac{\Lambda_i}{\id_{2k|2k}+\imath \Lambda_i \sigma_0}\delta\sigma\right)^2+(\sigma_0^{-1}\delta\sigma)^2-\imath \frac{2}{l}\tilde\xi\sigma_0\right]+\mathcal{O}\left(\frac{1}{p^{3/2}}\right).
 \end{split}
\end{align}
The term of first order in $\delta\sigma$ yields the supermatrix
valued saddle point equation
\begin{align}
 \label{Qeq}
\tilde{x}Q+  \id_{2k|2k} - \frac{\gamma^2}{p}\sum_{i=1}^{p} \frac{\Lambda_iQ}{\id_{2k|2k}+ \Lambda_iQ}=0 \ ,
\end{align}
where $Q$ denotes the solution.  We neglect the term in $\varepsilon$
as it is infinitesimal. The difficulty is that the saddle point
solution $Q=\imath\sigma$ depends in a highly non--trivial way on the
empirical eigenvalues $\Lambda_i$.

The saddle point equation is essentially scalar, as may be seen by
taking the commutator of Eq.~\eqref{Qeq} with $Q^{-1}$. We obtain
$\tilde x=Q^{-1}\tilde xQ$ implying that $Q$ and $\tilde x$
commute. Thus, we can analyze \eq{Qeq} in the space of the eigenvalues
of $Q$. There are two kinds of eigenvalues, namely $q^{({\rm
    b})}=\diag(q_1^{({\rm b})},\ldots, q_{2k}^{({\rm b})})$ in the
boson--boson block and $q^{({\rm f})}=\diag(q_1^{({\rm
    f})},\ldots,q_k^{({\rm f})})\otimes\id_2$ in the fermion--fermion
block. The double degeneracy of the latter is the Kramers degeneracy for
quaternion matrices. The integration domain is non--compact for $q^{\rm
  (b)}$, $q_j^{({\rm b})}\in \imath \tilde{L}_j\mathbb{R}_+$ with $\tilde{L}_j=\pm1$,
and compact for $q^{({\rm f})}$, $q_j^{({\rm f})}\in{\rm U}(1)$. Hence we
only need to analyze the scalar saddle point equation
\begin{align}
\label{ScalarSaddlePoint}
  0 = -x_a-\frac{1}{q(x_a)}+\frac{\gamma^2}{p}\sum_{i=1}^{p} \frac{\Lambda_i}
                           { 1 +  \Lambda_i q(x_a)}
   = -x_a+g(q(x_a)) = \hat{g}(q(x_a),x_a) \ ,
\end{align}
where we introduce the functions $q(x_a)$, $g(q(x_a))=\hat{g}(q(x_a),0)$ and
$\hat{g}(q(x_a),x_a)$.  The level density~\eqref{level-asymp} is directly related to the solutions of this
equation. Equation~\eqref{ScalarSaddlePoint} is a classical result in high dimensional
inference~\cite{MarcenkoPastur,silversteinBai,SilversteinChoi,Shinzato} where
it was derived by other means. Mar\u{c}enko and
Pastur~\cite{MarcenkoPastur} showed that, if this equation has a
solution in the upper half--plane, this solution is unique, which we
denote by $q_0(x_a)$. We briefly review the analysis of
the rational function $g(q)$ at finite matrix dimension $p$, in particular its singularities. We need results of this discussion for the analysis of the spectral statistics on the level of the local level spacing.

\begin{figure}
\begin{tikzpicture}[scale=0.6]
  \draw[ thick, red] [domain=0.225:3, samples=50] plot (\x, {-1-1/(1.5*\x) });
  \draw[ thick, red] [domain=-1.865:-0.135, samples=50] plot (\x, {-1/((2*\x)+(\x*\x))) });
  \draw[ thick, red] [domain=-2.38:-2.25, samples=50] plot (\x, {1/(2+\x) });
  \draw[ thick, red] [domain=-4.25:-4.12, samples=50] plot (\x, {1/(4.5+\x) });
  \draw[ thick, red] [domain=-4.88:-4.75, samples=50] plot (\x, {1/(4.5+\x) });
  \draw[ thick, red] [domain=-6.75:-6.62, samples=50] plot (\x, {1/(7+\x) });
  \draw[ thick, red] [domain=-10:-7.25, samples=50] plot (\x, {-1-0.7/(7+\x) -0.36/((7+\x)*(7+\x)) });
  %
  \draw[blue,very thick,->] (-10,0) -- (-7,0)  (-4.5,0) -- (3,0);
  \draw[blue,very thick]  (-7,0) --  (-4.5,0) node [midway,fill=white] {//};
  \draw[blue,very thick] (-9,0) -- (-7,0);
  \draw (2.8,-0.5) node {\Large $q$};
  \draw[blue,very thick,->] (0,-4) -- (0,3);
  \draw (0.9,3.5) node {\Large $\hat{g}(q)$};
  \draw[green, very thick, dashed] (-10,-1) -- (-7,-1);
  \draw[green, very thick, dashed] (-7,-1) -- (-4.5,-1) node [midway,fill=white] {//};
  \draw[green, very thick, dashed] (-4.5,-1) -- (3,-1);
  \node[draw,circle,inner sep=1.5pt,fill] at (0,-1) {};
  \draw (0.7,-1) node[below] { $-x$};
  \draw (-8.4,3.3) node {\Large $\gamma^2<1$};
  \draw[green, very thick, dashed] (0,-4) -- (0,4);
  \node[draw,circle,inner sep=1.5pt,fill] at (0,0) {};
  \draw (0.5,0) node[below] { $0$};
  \draw[green, very thick, dashed] (-2,-4) -- (-2,4);
  \node[draw,circle,inner sep=1.5pt,fill] at (-2,0) {};
  \draw (-1.9,0) node[below] { $-\Lambda_p^{-1}$};
  \draw[green, very thick, dashed] (-4.5,-4) -- (-4.5,4);  
  \node[draw,circle,inner sep=1.5pt,fill] at (-4.5,0) {};
  \draw (-4.27,0) node[below] { $-\Lambda_{p-1}^{-1}$};
  \draw[green, very thick, dashed] (-7,-4) -- (-7,4);  
  \node[draw,circle,inner sep=1.5pt,fill] at (-7,0) {};
  \draw (-6.9,0) node[below] { $-\Lambda_{1}^{-1}$};
\end{tikzpicture} \quad
\begin{tikzpicture}[scale=0.6]
  \draw[ thick, red] [domain=0.225:3, samples=50] plot (\x, {-1-1/(1.5*\x) });
  \draw[ thick, red] [domain=-1.865:-0.135, samples=50] plot (\x, {-1/((2*\x)+(\x*\x))) });
  \draw[ thick, red] [domain=-2.38:-2.25, samples=50] plot (\x, {1/(2+\x) });
  \draw[ thick, red] [domain=-4.25:-4.12, samples=50] plot (\x, {1/(4.5+\x) });
  \draw[ thick, red] [domain=-4.88:-4.75, samples=50] plot (\x, {1/(4.5+\x) });
  \draw[ thick, red] [domain=-6.75:-6.62, samples=50] plot (\x, {1/(7+\x) });
  \draw[ thick, red] [domain=-10:-7.25, samples=50] plot (\x, {-1+0.75/(7+\x) });
  %
  \draw[blue,very thick,->] (-10,0) -- (-7,0)  (-4.5,0) -- (3,0);
  \draw[blue,very thick]  (-7,0) --  (-4.5,0) node [midway,fill=white] {//};
  \draw[blue,very thick] (-9,0) -- (-7,0);
  \draw (2.8,-0.5) node {\Large $q$};
  \draw[blue,very thick,->] (0,-4) -- (0,3);
  \draw (0.9,3.5) node {\Large $\hat{g}(q)$};
  \draw[green, very thick, dashed] (-10,-1) -- (-7,-1);
  \draw[green, very thick, dashed] (-7,-1) -- (-4.5,-1) node [midway,fill=white] {//};
  \draw[green, very thick, dashed] (-4.5,-1) -- (3,-1);
  \node[draw,circle,inner sep=1.5pt,fill] at (0,-1) {};
  \draw (0.7,-1) node[below] { $-x$};
  \draw (-8.4,3.3) node {\Large $\gamma^2=1$};
  \draw[green, very thick, dashed] (0,-4) -- (0,4);
  \node[draw,circle,inner sep=1.5pt,fill] at (0,0) {};
  \draw (0.5,0) node[below] { $0$};
  \draw[green, very thick, dashed] (-2,-4) -- (-2,4);
  \node[draw,circle,inner sep=1.5pt,fill] at (-2,0) {};
  \draw (-1.9,0) node[below] { $-\Lambda_p^{-1}$};
  \draw[green, very thick, dashed] (-4.5,-4) -- (-4.5,4);  
  \node[draw,circle,inner sep=1.5pt,fill] at (-4.5,0) {};
  \draw (-4.27,0) node[below] { $-\Lambda_{p-1}^{-1}$};
  \draw[green, very thick, dashed] (-7,-4) -- (-7,4);  
  \node[draw,circle,inner sep=1.5pt,fill] at (-7,0) {};
  \draw (-6.9,0) node[below] { $-\Lambda_{1}^{-1}$};
\end{tikzpicture} 
\caption{(Color online) Asymptotic schematic behaviour of the rational
  function $\hat{g}(q,x)$ (solid curves) at the singularities (dashed) and at
  infinity, {\it cf.} Eq.~\eqref{ScalarSaddlePoint}. The variable $x$
  stands for any eigenvalue $x_1,\ldots,x_k$ of the correlated Wishart
  matrix $WW^T$. The qualitative convexity properties for
  $q>-1/\Lambda_p$ are also shown. However the behaviour for
  $q<-1/\Lambda_p$ strongly depends on the parameter $\gamma^2=p/n$
  and the empirical eigenvalues $\Lambda_a$. Concretely, we only have
  a maximum above the horizontal green line for $q<-1/\Lambda_1$ when
  $\gamma^2<1$ (left figure) and no maximum at all for $\gamma^2=1$
  (right figure). In the latter case $g$ approaches the value $-1$ for
  $q\to-\infty$ from below instead from above.  Moreover we may have
  maximally one maximum and one minimum inside any of the intervals
  $]-1/\Lambda_j,-1/\Lambda_{j+1}[$ or none at all, depending on the
  distance between the individual $\Lambda_a$.  }
\label{roots}
\end{figure}
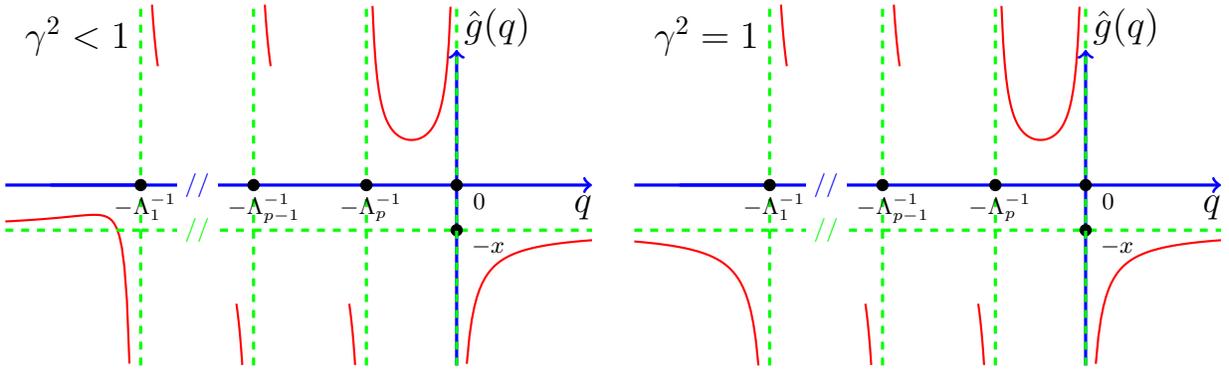

The equation $\hat{g}(q(x_a),x_a)=0$ has $p+1$ roots for each $x_a$. Moreover
the function $g(q)$ is singular at $q=-1/\Lambda_i$ for $i=1,\dots,p$
and at $q=0$. An asymptotic analysis of the singularities yields
\begin{align}\label{asymp-q-inf}
 \lim_{q\rightarrow0^\pm}g(q) = \mp \infty,\ \lim_{q\rightarrow(-1/\Lambda_i)^\pm}g(q) =\pm \infty\text{, and }\hat{g}(q(x_a),x_a)\overset{|q|\gg1}{\rightarrow} -x_a-\frac{1-\gamma^2}{q}-\sum_{i=1}^p\frac{\gamma^2}{p\Lambda_i}\frac{1}{q^2}
\end{align}
where $\pm$ indicates the limit from above or below, respectively. Figure~\ref{roots} shows the asymptotic behaviour of $\hat{g}(q,x)$.
As there is at least one real root of $g(q)$ within each interval
$\left(-1/\Lambda_{i+1},-1/\Lambda_{i}\right)$ for
$i=1,\dots,p-1$, at least $p-1$ out of $p+1$ roots are
real. Since \eq{ScalarSaddlePoint} is real, the complex conjugate
$q_0^*(x_a)$ of a solution $q_0(x_a)$ solves \eq{ScalarSaddlePoint} as
well. Hence, the remaining two roots are either a complex conjugate
pair or both real.
 
When a complex conjugate pair solves the saddle point equation, the
eigenvalues $q_j^{({\rm b})}$ in the boson--boson block of the
supermatrix $Q$ can only reach those solutions which
share the same sign of the imaginary part with $x_a+\imath
L_a\varepsilon$. This is due to the infinitely high potential walls around the
singularities $q=-1/\Lambda_i$ when $p\to\infty$. In contrast, the
eigenvalues $q_j^{({\rm f})}$ in the fermion--fermion block reach both
saddle points.  When diagonalizing the supermatrix $Q$ we obtain the
Berezinian, \textit{i.e.}, the superspace Jacobian,
\begin{equation}\label{Ber}
 B_{2k|k}(q^{({\rm b})};q^{({\rm f})})=\frac{|\Delta_{2k}(q^{({\rm b})})|^3\Delta_k^6(q^{({\rm f})})}
                                     {\Delta_{3k}^2(q^{\rm (b)};q^{({\rm f})})}
\end{equation}
with the Vandermonde determinant $\Delta_k(y)=\prod_{1\leq
  i<j\leq k}(y_j-y_i)$. We plug the two kinds
of saddle points into this Berezinian. Solutions in which the eigenvalues of the boson--boson block
and the fermion--fermion block do not agree are algebraically
suppressed by factors of $1/p$ and thus smaller than those in which the
spectra of the boson--boson and the fermion--fermion blocks
counted with multiplicities coincide.

In the case that all solutions are real we may reach more than one
saddle point with the boson--boson block of $Q$. However only one of
all real saddle points contributes because of the particular behavior of the extrema of $g$. Hence, we have to consider the
first and second derivative of $g$ (second and third derivative of the
Lagrangian~\eqref{approxLagrangian}) which read
\begin{align}\label{g-der}
  g'(q)= \frac{1}{q^2}-\frac{\gamma^2}{p}\sum_{i=1}^{p} \frac{\Lambda_i^2}{ (1 +  \Lambda_i q)^2},\quad g''(q)=- \frac{3}{q^3}+\frac{3\gamma^2}{p}\sum_{i=1}^{p} \frac{\Lambda_i^3}{ (1 +  \Lambda_i q)^3}.
\end{align}
For $q\in]-1/\Lambda_p,0[$ the function $g$ is concave and for $q>0$ it is convex due to the estimates
\begin{align}
  g''(q)\geq 3\gamma^2\lambda_p^3-\frac{3}{q^3}\geq 3(\gamma^2+1)\Lambda_p^3>0\text{ and }g''(q)\leq 3\frac{\gamma^2-1}{q^3}\leq 0
\end{align}
since $\gamma^2\leq 1$ and $(q+1/\Lambda_j)^{-3}\geq \Lambda_j^3$ for all
$j=1,\ldots,p$. Hence there is only one minimum in $]-1/\Lambda_p,0[$, {\it cf.}
Fig.~\ref{roots}. When $q$ is between two empirical eigenvalues, in particular
$q\in]-1/\Lambda_j,-1/\Lambda_{j+1}[$ with $j=1,\ldots,p-1$, we find
either not an extremum or a single pair of a minimum and a
maximum. This results from the curvature of $q'$ which is the third derivative of $q$,
\begin{align}
  \frac{g^{(3)}(q)}{6}&=\frac{1}{q^4}-\frac{\gamma^2}{p}\sum_{i=1}^{p} \frac{\Lambda_i^4}{ (1 +  \Lambda_i q)^4}
  \leq \max\left\{0,\left(\frac{1}{p}\sum_{i=1}^{p} \frac{\Lambda_i^2}{ (1 +  \Lambda_i q)^2}\right)^2-\frac{1}{p}\sum_{i=1}^{p} \frac{\Lambda_i^4}{ (1 +  \Lambda_i q)^4}\right\}\leq0\label{est-conv}
\end{align}
for all $q$ satisfying $g'(q)<0$. Hence $g'(q)$ is convex in these regimes
implying the extrema for $g(q)$. In Eq.~\eqref{est-conv} we employed the assumption $g'(q)<0$, the fact
that the right hand side is a concave function in $\gamma^2\in[0,1]$
and that the second term in the maximization is the negative variance
of the sequence $\Lambda_i^2/(1 + \Lambda_i q)^2$, $i=1,\ldots,p$. The
estimate~\eqref{est-conv} also tells us that for $q<-1/\Lambda_1$ the
function has either a single maximum or none at all depending on
whether $\gamma^2<1$ or $\gamma^2=1$, respectively.
 
Only those solutions $q_0(x)$ where $g(q_0(x))$ has a positive slope along the contour
correspond to a minimum of the Lagrangian~\eqref{approxLagrangian} in
the eigenvalues of the boson--boson block of $-Q\tilde x^{-1}=-\tilde
L \sigma\tilde x^{-1}$. The asymptotic behaviour and the convexity
properties of the rational function $g$ imply that in the case of
$p+1$ real solutions only one of those solutions has a positive slope
at $g(q_0(x))$. Consequently, the non--compact integrals are evaluated
at this saddle point only regardless what sign $L_a$ is chosen. A
similar argument holds for the compact integrals over $q_{\rm f}$ in
the fermion-fermion block which sees exactly the same point as a
minimum as the eigenvalues $q_{\rm f}$ and all other real solutions
appear as maxima in the Lagrangian~\eqref{approxLagrangian}. Despite
the fact that the contours of $q^{({\rm b})}$ and $q^{({\rm f})}$
orthogonally cross each other at the saddle point $q_0(x)$, the
opposite sign in the supertrace renders the saddle point for both
contours a minimum.  Summing over $L$ in \eq{Rk} we notice that those
terms where the contributing saddle point solution is real vanish
because the contributing saddle point is independent of the
corresponding sign. Therefore only the complex solutions
contribute. We thus omit all real roots in the following.

Since $Q$ and $\tilde x$ commute, we may choose an appropriate block
diagonal basis, $Q=\diag(Q^{(1)},\dots,Q^{(\alpha)})$ where $\alpha<k$
is the number of distinct points $x_a$, and discuss the resulting
saddle point equation for each block separately. The size of a single
block depends on the degeneracy $m_o$ of the point $x^{(o)}$ in
question, $o=1,\ldots,\alpha$, \textit{i.e.} if
$x_{i_1}=x_{i_2}=\dots=x_{i_{m_o}}$ the corresponding block is of
dimension $(2m_o|2m_o)$ in superspace. By $\tilde
L^{(o)}=L^{(o)}\otimes\id_{1|1}\otimes\id_2$ we denote the projection
of $\tilde L$ onto the block corresponding to the point $x^{(o)}$. The
resulting saddle point equation is invariant with respect to
$\text{UOSp}(\tilde L^{(1)})\times\cdots\times\text{UOSp}(\tilde
L^{(\alpha)})$, where $\text{UOSp}(\tilde L^{(o)})$ is the group of
pseudo--unitary orthosymplectic matrices $T$ with the property
$T\diag(L^{(o)}\otimes\id_2;\id_{2m_o}) T^\dagger =
\diag(L^{(o)}\otimes\id_2;\id_{2m_o})$. Hence, instead of isolated
saddle points we obtain saddle point manifolds, see
\Refs{Verbaarschotetsal,efetov,Zirnbauer1996,GuhrSUSY,SUSYZirnbauer}
in another context. From the discussion above, we have to integrate
$Q=\imath\sigma_0=\diag(Q^{(1)},\dots,Q^{(\alpha)})$ with
\begin{align}
 Q^{(o)} =& \text{Re}~q_0(x^{(o)})\id_{m_o} +\imath \, T^{(o)}  \text{Im}~q_0\left(x^{(o)}+\imath \tilde L^{(o)}\varepsilon\right) {T^{(o)}}^{-1}
 \approx \text{Re}~q_0(x^{(o)})\id_{m_o} +\imath\text{Im}~q_0(x^{(o)})\, T^{(o)}  \tilde L^{(o)} {T^{(o)}}^{-1}
\end{align}
over the coset
\begin{align}
 T^{(o)}\in \text{UOSp}(\tilde L^{(o)})/[\text{UOSp}(2m_0^{(o)}|2m_0^{(o)})\times\text{UOSp}(2m_1^{(o)}|2m_1^{(o)})]
\end{align}
which parametrizes the ``Goldstone modes''.  The variables
$2m_0^{(o)}$ and $2m_1^{(o)}$ are the numbers of $+1$'s and $-1$'s in
$\tilde L^{(o)}$ such that $m_0^{(o)}+m_1^{(o)}=m_o$.

We now turn to the integration over the ``massive modes'', parametrized by
\begin{equation}
\delta\sigma=T\left[\begin{array}{ccc} \delta\sigma_{11} & \cdots & \delta\sigma_{1\alpha} \\ \vdots & & \vdots \\ \delta\sigma_{\alpha 1} & \cdots & \delta\sigma_{\alpha\alpha} \end{array}\right]T^{-1}
\end{equation}
where $T=\diag(T^{(1)},\ldots,T^{(\alpha)})$ and where
$\delta\sigma_{ab}$ is a $(2m_a|2m_a)\times(2m_b|2m_b)$
supermatrix. The diagonal blocks satisfy the commutation relations
$[\delta\sigma_{oo},\tilde L^{(o)}]=0$ since the remaining
integration, in particular the components which do not commute, is
accounted for by the integrals over $T^{(o)}$. The challenging part in
specifying the whole symmetries of the blocks $\delta\sigma_{ab}$ are
the phases in front. The quadratic part in $\delta\sigma$ of the
Lagrangian~\eqref{approxLagrangian.b} has to be positive definite and
must ensure convergence. We define the complex numbers
\begin{equation}
 z_{ab}^{(+)}=\frac{\gamma^2}{p}\sum_{i=1}^p\frac{1}{\Lambda_i^{-1}+q_0(x^{(a)})}\frac{1}{\Lambda_i^{-1}+q_0(x^{(b)})}-\frac{1}{q_0(x^{(a)})q_0(x^{(b)})}
\end{equation}
and
\begin{equation}
 z_{ab}^{(-)}=\frac{\gamma^2}{p}\sum_{i=1}^p\frac{1}{\Lambda_i^{-1}+q_0(x^{(a)})}\frac{1}{\Lambda_i^{-1}+(q_0(x^{(b)}))^*}-\frac{1}{q_0(x^{(a)})(q_0(x^{(b)}))^*} \ ,
\end{equation}
allowing us to split the matrix blocks as follows
\begin{eqnarray}
\delta\sigma_{aa}&=&\frac{1}{\sqrt{z_{aa}^{(+)}}}\delta\sigma_{aa}^{(00)}+\frac{1}{\sqrt{(z_{aa}^{(+)})^*}}\delta\sigma_{aa}^{(11)},\label{sigma:aa}\\
\delta\sigma_{ab}&\overset{a<b}{=}&\frac{1}{\sqrt{z_{ab}^{(+)}}}\delta\sigma_{ab}^{(00)}+\frac{1}{\sqrt{(z_{ab}^{(+)})^*}}\delta\sigma_{ab}^{(11)}+\frac{1}{\sqrt{z_{ab}^{(-)}}}\delta\sigma_{ab}^{(01)}+\frac{1}{\sqrt{(z_{ab}^{(-)})^*}}\delta\sigma_{ab}^{(10)},\label{sigma:ab}\\
\delta\sigma_{ab}&\overset{a>b}{=}&\frac{1}{\sqrt{z_{ab}^{(+)}}}{\delta\sigma_{ab}^{(00)}}^\dagger+\frac{1}{\sqrt{(z_{ab}^{(+)})^*}}{\delta\sigma_{ab}^{(11)}}^\dagger+\frac{1}{\sqrt{z_{ab}^{(-)}}}{\delta\sigma_{ab}^{(01)}}^\dagger+\frac{1}{\sqrt{(z_{ab}^{(-)})^*}}{\delta\sigma_{ab}^{(10)}}^\dagger,\label{sigma:ac}
\end{eqnarray}
such that $\tilde L^{(i)} \delta\sigma_{ab}^{(ij)}\tilde
L^{(j)}=(-1)^{i+j}\delta\sigma_{ab}^{(ij)}$. Blocks of the form
$\delta\sigma_{aa}^{(01)}$ and $\delta\sigma_{aa}^{(10)}$ do not
exist, because of the required commutation relation with $\tilde
L^{(a)}$. We recall the dimensions $m_0^{(a)}$ and $m_1^{(a)}$ which essentially are
the signature of $\tilde L^{(a)}$. The diagonal matrix blocks
$\delta\sigma_{aa}^{(jj)}$ are Hermitian supermatrices of dimension
$(2m_j^{(a)}|2m_j^{(a)})\times(2m_j^{(a)}|2m_j^{(a)})$ where the
boson--boson blocks are real symmetric and the fermion--fermion blocks
are Hermitian self--dual. The off--diagonal block
$\delta\sigma_{ab}^{(ij)}$ has dimension
$(2m_i^{(a)}|2m_i^{(a)})\times(2m_j^{(b)}|2m_j^{(b)})$. Its
boson--boson block is an arbitrary real matrix and its fermion-fermion
block an arbitrary quaternion matrix. The integration measure of $T$
is the Haar measure on the coset and the one of $\delta\sigma$ is the
flat Lebesgue measure for the commuting and the Berezin measure for
the anticommuting variables. Collecting everything, the Lagrangian~\eqref{approxLagrangian.b}
takes the form
\begin{align}
\label{approxLagrangian.c}
 \begin{split}
\mathcal L \left(\sigma_0+\frac{\delta\sigma}{\sqrt{p}}\right) =&  -\imath\sum_{o=1}^{\alpha}{\rm Im}\, q_0(x^{(o)})\Str\left(\imath\varepsilon\tilde L^{(o)}+\frac{\tilde\xi^{(o)}}{pl}\right)  T^{(o)}  \tilde L^{(o)} {T^{(o)}}^{-1} \\
&+\frac{1}{p}\sum_{1\leq a<b\leq\alpha}\sum_{i,j=0,1}\Str\delta\sigma_{ab}^{(ij)}{\delta\sigma_{ab}^{(ij)}}^\dagger+\frac{1}{2p}\sum_{a=1}^\alpha\sum_{j=0,1}\Str(\delta\sigma_{aa}^{(jj)})^2+\mathcal{O}\left(\frac{1}{p^{3/2}}\right)
 \end{split}
\end{align}
where $\tilde\xi=\diag(\tilde\xi^{(1)},\ldots,\tilde\xi^{(\alpha)})$
is splitted analogously to $\tilde{L}$. The prefactors of the
individual blocks $\delta\sigma_{ab}^{(ij)}$, see Eqs.~(\ref{sigma:aa}-\ref{sigma:ac}), cancel in the
Berezinian after the change of coordinates for the supermatrix $\sigma$ into the Goldstone and the massive modes
because we have for each of these blocks the same number of real
variables and Grassmann variables.

To proceed we have to carefully analyze the Efetov--Wegner boundary
terms~\cite{efetov,Wegner}. They are an inherent feature of
superanalysis without counterpart in ordinary analysis. These terms
appear whenever a change of variables is performed on superspaces with boundaries, including those boundaries induced by coordinate singularities of the Berezinian.

\subsection{Macroscopic Level Density}\label{sec:Level}

In the case of the macroscopic level density, \textit{i.e.}, $k=1$, $\xi_1=0$
and $\tilde{L}^{(1)}=L_1^{(1)}\id_{2|2}$,
Efetov--Wegner boundary terms cannot appear because we only shift and rescale
the supermatrix $\sigma\to\delta\sigma$. The 
Gaussian integral over $\delta\sigma$ cancels the constant
$\lim_{n\to\infty}K_{nl,1}=1/(8\pi^2)$ in the limit $n\to\infty$. The level density becomes
\begin{align}\label{level-asymp}
R_1(x) \overset{n\gg1}{\approx}\lim_{\varepsilon\rightarrow0}\text{Im}\left(\frac{1}{\gamma^2\pi }q_0(x+\imath\varepsilon)+\frac{\gamma^{-2}-1}{\pi}\frac{1}{x+\imath \varepsilon}\right)=\frac{1}{\gamma^2\pi }\text{Im}~q_0(x),
\end{align}
for all values of $l$. Hence, the saddle point solution $q_0(x)$
is up to the normalization $1/\gamma^2$ the Green function --- also
known as Cauchy or Stieltjes transform --- of the density $R_1(x)$.
When writing $\text{Im}~q_0(x)$ shorthand, we view the $1/x$
singularity of $q_0(x)$ at the origin as a real term which may be
neglected. Thus, the chain of equalities~\eqref{level-asymp} is
consistent.

The coincidence of $q_0(x)/\gamma^2$ with the Green function
implies that the function $g(q)-1/q$, see
Eq.~\eqref{ScalarSaddlePoint}, can be identified with the R transform
in the theory of free probability.  An introduction to free
probability in random matrix theory can be found in
Ref.~\cite{Burda,Speicher}. Free probability in the context of random matrices was originally introduced by Voiculescu et al.~\cite{Voiculescu}.

The Dirac $\delta$ function or equivalently the second term under the
limit in Eq.~\eqref{level-asymp} is important for $\gamma^2<1$ when
the limit $\varepsilon\to0$ is still to be taken. To clarify this we
consider the asymptotics of the saddle point solution for
$x+\imath\varepsilon\to0$ which is equivalent to $q\to-\infty$,
\textit{cf.}~Fig.~\ref{roots}. We employ the
asymptotics~\eqref{asymp-q-inf} of the function $\hat{g}$. Taking into
account only the first two terms, we find the asymptotic behaviour of
the saddle point solution as
$q_0(x+\imath\varepsilon)\approx(\gamma^2-1)/(x+\imath\varepsilon)$
for $|x+\imath\varepsilon|\ll1$. The imaginary part of this term
yields in the limit $\varepsilon\to0$ the Dirac $\delta$ function at
the origin which we subtract.

To study the edges of the spectral support we again start from
the saddle point equation~\eqref{ScalarSaddlePoint}. Multiplying this
equation with $q_0(x)$ and taking the imaginary part for $x>0$ we find
\begin{equation}\label{Im-saddle}
x\,{\rm Im}\, q_0(x)=\frac{\gamma^2}{p}\sum_{j=1}^p\frac{\Lambda_j{\rm Im}\,q_0(x)}{(1+\Lambda_j {\rm Re}\,q_0(x))^2+\Lambda_j^2({\rm Im}\,q_0(x))^2}.
\end{equation}
A similar equation can be derived for the real part,
\begin{equation}\label{Re-saddle-a}
x\,{\rm Re}\, q_0(x)=\gamma^2-1-\frac{\gamma^2}{p}\sum_{j=1}^p\frac{1+\Lambda_j{\rm Re}\,q_0(x)}{(1+\Lambda_j {\rm Re}\,q_0(x))^2+\Lambda_j^2({\rm Im}\,q_0(x))^2}.
\end{equation}
The latter equation can be rewritten to
\begin{equation}\label{Re-saddle-b}
{\rm Re}\, q_0(x)=\frac{\gamma^2-1-\gamma^2/p\sum_{j=1}^p1/[(1+\Lambda_j {\rm Re}\,q_0(x))^2+\Lambda_j^2({\rm Im}\,q_0(x))^2]}{x+\gamma^2/p\sum_{j=1}^p\Lambda_j/[(1+\Lambda_j {\rm Re}\,q_0(x))^2+\Lambda_j^2({\rm Im}\,q_0(x))^2]}<0
\end{equation}
which is obviously always negative because $\gamma^2=p/n\leq1$. Hence
the sum $1+\Lambda_{j_0}{\rm Re}\,q_0(x)$ might vanish for a
particular $\Lambda_{j_0}$ such that we have to be careful. However
this scenario does not happen at an edge where either ${\rm Im}\,
q_0(x)\to0$ or ${\rm Im}\, q_0(x)\to\infty$ due to the
following reason. Suppose ${\rm Re}\,q_0(x)=-\Lambda_{j_0}^{-1}$ and
$\Lambda_{j_0}$ has the degeneracy $l_0$, Eq.~\eqref{Im-saddle}
reads
\begin{equation}\label{Im-saddle-b}
x\,{\rm Im}\, q_0(x)=\frac{\gamma^2}{p}\sum_{\Lambda_j\neq\Lambda_{j_0}}\frac{\Lambda_j{\rm Im}\,q_0(x)}{(1-\Lambda_j /\Lambda_{j_0})^2+\Lambda_j^2({\rm Im}\,q_0(x))^2}+\frac{\gamma^2 l_0}{p}\frac{1}{\Lambda_{j_0}{\rm Im}\,q_0(x)}
\end{equation}
which is never satisfied by one of the two solutions ${\rm
  Im}\,q_0(x)=0,\infty$. Thus we only have $1+\Lambda_{j}{\rm
  Re}\,q_0(x)\neq0$ for all $j=1,\ldots,p$ at an edge.

For Eq.~\eqref{Im-saddle}, there are only two types of
solutions. Either we are at the origin, then ${\rm Im}\,q_0(x)$ has to
diverge, according to ${\rm Im}\,q_0(x)=c/\sqrt{x}$ with
$c^{-1}=\gamma^2/p\sum_{j=1}^p1/\Lambda_j$, to satisfy
Eq.~\eqref{Im-saddle}, or the edge is not at the origin, then we can
expand Eq.~\eqref{Im-saddle} for small ${\rm Im}\,q_0(x)$ which yields
the square root behavior
\begin{equation}\label{asymp-edge-saddle}
{\rm Im}\,q_0(x) \sim \left\{\begin{array}{cl} \sqrt{x-x_{\rm edge}}, & x_{\rm edge}\text{ is a lower bound of a cut},\\  \sqrt{x_{\rm edge}-x}, & x_{\rm edge}\text{ is an upper bound of a cut}. \end{array}\right.
\end{equation}
The largest and smallest eigenvalue lie at the edges
\begin{equation}
x_{\max}=g\left(-\int\limits_{-\Lambda_p^{-1}}^0\Theta(g'(q)){\rm d}q\right)\text{ and }x_{\min}=g\left(-\int\limits_{-\infty}^{-\Lambda_1^{-1}}\Theta(-g'(q)){\rm d}q-\Lambda_1^{-1}\right)
\end{equation}
with the Heaviside function $\Theta$. This result follows
from the saddle point equation~\eqref{ScalarSaddlePoint} and from the
monotonic behavior of $g'(x)$,
\textit{cf.}~Fig.~\ref{roots}.
When we have more than only one cut in the spectrum, we find upper edges at
\begin{equation}
x_{\rm u}^{(j)}=g\left(\left[\left(\int\limits_{-\Lambda_{j-1}^{-1}}^{-\Lambda_{j}^{-1}}\Theta(g'(q)){\rm d}q\right)^2-2\int\limits_{-\Lambda_{j-1}^{-1}}^{-\Lambda_{j}^{-1}}q\Theta(g'(q)){\rm d}q\right]\biggl/\left[2\int\limits_{-\Lambda_{j-1}^{-1}}^{-\Lambda_{j}^{-1}}\Theta(g'(q)){\rm d}q\right]\right)
\end{equation}
and lower edges at
\begin{equation}
x_{\rm l}^{(j)}=g\left(\left[\left(\int\limits_{-\Lambda_{j-1}^{-1}}^{-\Lambda_{j}^{-1}}\Theta(g'(q)){\rm d}q\right)^2+2\int\limits_{-\Lambda_{j-1}^{-1}}^{-\Lambda_{j}^{-1}}q\Theta(g'(q)){\rm d}q\right]\biggl/\left[2\int\limits_{-\Lambda_{j-1}^{-1}}^{-\Lambda_{j}^{-1}}\Theta(g'(q)){\rm d}q\right]\right)
\end{equation}
in the interval $]\Lambda_{j-1},\Lambda_{j}[$ with
$j=2,\ldots,p$. Edges are not found in $]\Lambda_{j-1},\Lambda_{j}[$
when $x_{\rm u}^{(j)}=x_{\rm l}^{(j)}$, in particular as
$g'(q)$ is strictly negative in
$]-\Lambda_{j-1}^{-1},-\Lambda_{j}^{-1}[$. It might happen that
two cuts start merging such that the latter scenario occurs, too. Then
one has to take into account the second derivative $g''(q)$ and the
level density behaves as $(x_{\rm edge}-x)^{1/3}$ where we expect
Pearcy kernel~\cite{Pearcy-org,Pearcy} behavior on the local
scale. We do not show this in the present work.

The situation slightly changes when considering the exact limit
$n,p\to\infty$ and $\gamma^2=p/n$ fixed where we have to assume a
limiting density $\rho(\lambda)$ for the empirical eigenvalues
$\Lambda$, {\it cf.} Eq.~\eqref{lim-dis.a}. As long as $-{\rm Re}\,
q_0(x)$ is in the support of the empirical density $\rho(\lambda)$
where this density is finite we can carry out the same analysis as
above because the saddle point solution has to satisfy the counterpart
of Eq.~\eqref{Im-saddle} which is
\begin{equation}\label{saddle-Im-int}
 x{\rm Im}\,q_0(x)=\gamma^2\int\limits_0^\infty \frac{\lambda\rho(\lambda){\rm d}\lambda}{(1+\lambda{\rm Re}\,q_0(x))^2+\lambda^2({\rm Im}\,q_0(x))^2}{\rm Im}\,q_0(x).
\end{equation}
As the integrand is divergent for ${\rm Im}\,
q_0(x)=0$ we conclude that ${\rm Im}\, q_0(x)$ has to be finite. This
argument also applies when $-{\rm Re}\, q_0(x)$ is at an edge of
$\rho(\lambda)$ where the density either diverges (this
divergence has to be integrable and to satisfy
assumption~\eqref{assumption}) or remains finite. Here we exclude the
origin where the behavior is different.

When $-{\rm Re}\, q_0(x)$ is taken at an edge where $\rho(\lambda)$
vanishes it may happen that ${\rm Im}\, q_0(x)$ vanishes, too, which
is, however, very unlikely. In particular we would expect this
scenario only when cuts may start to merge implying that the edge is
located in the bulk of the spectrum. The generic case is that ${\rm
  Im}\, q_0(x)$ vanishes when $-{\rm Re}\, q_0(x)$ is outside of the
support of the empirical density $\rho(\lambda)$. Hence, if this is
the case and we are at a soft edge, {\it i.e.} ${\rm Im}\,
q_0(x)\to0$, we may expand Eq.~\eqref{saddle-Im-int} for small ${\rm
  Im}\, q_0(x)$ and find the square root
behavior~\eqref{asymp-edge-saddle}.

A hard edge (with ${\rm Im}\, q_0(x)\to\infty$) of the macroscopic
level density~\eqref{level-asymp} only appears at the origin
$x=0$. This follows from Eq.~\eqref{saddle-Im-int} when
$\gamma^2=p/n\to1$. We find the standard $1/\sqrt{x}$ behavior in the
case that $\rho(\lambda)$ is separated by a finite gap from the
origin. The situation drastically changes when the support of
$\rho(\lambda)$ touches the origin. For example for
$\rho(\lambda)=\Theta(1-\lambda)$ we find a singular behavior with
$\sqrt{{\rm ln}\, x/x}$. The condition for encountering the standard
singularity $1/\sqrt{x}$ is the existence of the integral
$\int_0^\infty \rho(\lambda)d\lambda/\lambda$.

We restrict ourselves to a detailed discussion of the soft edges
having the form~\eqref{asymp-edge-saddle} in
section~\ref{sec:saddlepoint-edge}. On the local scale,
we will find the Airy statistics as for the uncorrelated Wishart
ensemble.

\subsection{Correlation Functions}\label{sec:k-point}

We turn to the $k$--point correlations for arbitrary $k\in\mathbb{N}$. We may
assume that $x^{(o)}>0, \ o=1,\ldots,\alpha$ and that these points do
not lie at a boundary of the support of the spectral
density~\eqref{level-asymp}. We thus omit the Dirac $\delta$
contributions at the origin, in particular the terms $1/(x_j+\imath
L_j \varepsilon)$ in Eq.~\eqref{eq:dd:kpointfunction}. We integrate
over the non--diagonal supermatrix blocks $\delta\sigma_{ab}$ ($a\neq
b$) which yields a constant equal to $(2\pi^2)^{2m_am_b}$ for the
block $\delta\sigma_{ab}$. We recall that Efetov--Wegner terms do not
occur since we only rescale those blocks.

The remaining integrations produce the well--known spectral statistics
built upon the sine kernel for real eigenvalues.  To show this, we
recall --- in an appropriate formulation --- the integral
representation of the $k$--point correlation functions on the local
scale of a Gaussian Orthogonal Ensemble (GOE) of $nl \times nl$
real symmetric matrices $H$,
 see \cite{Zirnbauer1996}
\begin{align}
X_{k}(\widehat\xi) =&\lim_{\substack{p\to\infty\\ \varepsilon\to0 \\ j\to0}}\sum_{L\in\{\pm 1\}^k}\prod_{i=1}^k \frac{L_i}{4 \imath nl} \partial_{j_i}\frac{\int\leb{H}\exp(-nl\,\tr H^2)\Sdet^{-1/2}(H\otimes\id_{2k|2k}-\id_{nl}\otimes(\pi\widehat{\xi}/(nl)+\tilde j+\imath \varepsilon\tilde L))}{\int\leb{H}\exp[-nl\,\tr H^2]}\nonumber\\
=&\lim_{\substack{n\to\infty\\ \varepsilon\to0}}K_{n}\sum_{L\in\{\pm 1\}^k}\int \leb{\sigma}\exp\left(-\frac{nl}{2}\widehat{\mathcal L}(\sigma)\right)\prod_{i=1}^k\frac{L_j }{8}\str\sigma\left[\begin{array}{cc} e^k_{jj}&0\\0&- e^k_{jj}\end{array}\right]\otimes\id_{2}\label{GOE-k-point}
\end{align}
where $\sigma$ is a supermatrix with the same symmetries as in Eq.~\eqref{superZ}. The Lagrangian is given by
\begin{equation}
\widehat{\mathcal L}(\sigma)=\frac{1}{2}\str\sigma^2-\imath\str\left(\imath\varepsilon\tilde{L}+\frac{\pi\widehat{\xi}}{nl}\right)\sigma-\left(1-\frac{1}{nl}\right)\str{\rm ln}\,\sigma.\label{Lag-GOE}
\end{equation}
The scaling of the local fluctuations $\pi\widehat{\xi}/(nl)$ is
motivated by the local GOE level spacing at the origin.  The
$(2k|2k)\times(2k|2k)$ supermatrix $\sigma$ is integrated over the
same domain as in Eq.~\eqref{superZ}. This $k$--point correlation
function~\eqref{GOE-k-point} contains the real sine kernel as can also
be derived by other methods such as skew--orthogonal polynomials,
\textit{e.g.}, see Refs.~\cite{MehtasBook,skew-2,skew-1}.

The saddle point equation in the limit $n\to\infty$ of
Eq.~\eqref{GOE-k-point} is simply $\sigma=\sigma^{-1}$. After an
analysis similar to the one in subsection~\ref{saddlepoint} we find
the saddle point manifold $\sigma=T(\tilde L
+\delta\sigma/\sqrt{n})T^{-1}$ with $T\in{\rm UOSp}(\tilde L)/[{\rm
  UOSp}(2k_0|2k_0)\times{\rm UOSp}(2k_1|2k_1)]$ where $\delta\sigma$
is any $\delta\sigma_{aa}$ in Eq.~\eqref{sigma:aa} with $m^{(a)}$
replaced by $k$. The integers $k_0$ and $k_1$ are the numbers of $+1$'s and $-1$'s of the $L_j$'s, respectively. The Lagrangian~\eqref{Lag-GOE} becomes
\begin{equation}
\widehat{\mathcal L}\left(T\tilde L T^{-1}+\frac{\delta\sigma}{\sqrt{n}}\right)=\frac{1}{2n}\str\delta\sigma^2-\imath\str\left(\imath\varepsilon\tilde{L}+\frac{\widehat{\xi}}{nl}\right)T\tilde L T^{-1}+\mathcal{O}\left(\frac{1}{n^{3/2}}\right)\label{Lag-GOE-approx}
\end{equation}
which we compare with the approximation~\eqref{approxLagrangian.c} of
the Lagrangian for the correlated Wishart model. Thus, the
identification
\begin{align}\label{unfolding}
\widehat{\xi}_a^{(o)} =R_1(x^{(o)})\xi_a^{(o)} \quad \Rightarrow \quad \dd\widehat{\xi}_a^{(o)} =R_1(x^{(o)})\dd\xi^{(o)}
\end{align}
with $k\to m^{(a)}$ must yield the same approximation. Indeed,
Eq.~\eqref{unfolding} is the unfolding prescription to uncover the
local spectral fluctuations $\xi_a^{(o)}$ at the position $x^{(o)}>0$.
 
To further solidify our line of reasoning, we now show that the
remaining parts of the integrand~\eqref{eq:dd:kpointfunction-asymp}
agree with this unfolding. Abbreviating the Efetov--Wegner boundary terms
with ``${\rm b.t.}$", we have
\begin{align}
 \label{eq:dd:kpointfunction-asymp}
 \begin{split}
 \lim_{p\to\infty} R_k(x,\xi)\leb{\xi} =& \prod_{o=1}^\alpha \biggl[\frac{K_{\infty,m_o}\leb{\xi^{(o)}}}{K_{\infty,m_0^{(o)}}K_{\infty,m_1^{(o)}}}\lim_{\varepsilon\rightarrow0}\sum_{L_1^{(o)},\ldots,L_{m_o}^{(o)}=\pm 1}
 \int {\rm d}\mu(T^{(o)})\left(\prod_{j=1}^{m_o}\frac{ L_j^{(o)}{\rm Im}\,q_0(x^{(o)})}{8\pi \gamma^2}\right.\\
 &\hspace*{-3cm}\times\left. \str T^{(o)}\tilde L^{(o)}{T^{(o)}}^{-1}\left[\begin{array}{cc} e^{m_o}_{jj}&0\\0&- e^{m_o}_{jj}\end{array}\right]\otimes\id_{2}\right)\exp\left(\frac{\imath{\rm Im}\, q_0(x^{(o)})}{2\gamma^2}\Str\tilde\xi^{(o)}T^{(o)}  \tilde L^{(o)} {T^{(o)}}^{-1}-\varepsilon\Str T^{(o)}\tilde L^{(o)}{T^{(o)}}^{-1}\tilde L^{(o)}\right)\biggl]+{\rm b.t.}\\
 =&\prod_{o=1}^\alpha X_{m^{(o)}}(\widehat\xi_1^{(o)},\dots,\widehat\xi_{m^{(o)}}^{(o)})\leb{\widehat\xi^{(o)}}.
 \end{split}
\end{align}
The ratio of the constants $K_{\infty,m_o}$, see
Eq.~\eqref{asymp-const}, in front of the flat measure
$\leb{\xi^{(o)}}$ results from the original constant $K_{nl,k}$ and
from the integration over $\delta\sigma$.  The real parts of
$q_0(x^{(o)})$ drop out because the corresponding integrands are
symmetric under the transformation $T^{(o)}\to VT^{(o)}$ where $V$
embeds the supergroup ${\rm UOSp}(2|2)$. This embedding in the form of
a $(2|2)\times(2|2)$ supermatrix corresponds to the diagonal matrix
$\diag(e^{m_o}_{jj},e^{m_o}_{jj};-e^{m_o}_{jj},-e^{m_o}_{jj})$ which
breaks this symmetry for the imaginary parts of
$q_0(x^{(o)})$. Adjusting Cauchy--like integration theorems
\textit{\`a la} Wegner~\cite{PS1979,efetov,Wegner, Con, ConGro,KKG08}
to our case of ${\rm UOSp}(2|2)$ we find that the corresponding blocks
of $T^{(o)}$ vanish such that the sign $L_j$ drops out in the
integrand, including the Lagrangian, and the sum over $L_j$ cancels
this contribution. Hence the integral only depends on the imaginary
part of the saddle point.

The measure ${\rm d}\mu(T^{(o)})$ is the Haar measure on the coset
$\text{UOSp}(\tilde
L^{(o)})/[\text{UOSp}(2m_0^{(o)}|2m_0^{(o)})\times\text{UOSp}(2m_1^{(o)}|2m_1^{(o)})]$. Its
normalization is induced by the flat measure $\leb{\sigma}$ from which
we started.  The $\varepsilon$ term in the exponential function still
guarantees absolute convergence of the integral because we may have
non--compact group integrals comprised in $T^{(o)}$. We absorbed the
prefactor in this latter term since it is a rescaling of $\varepsilon$ and we take the limit
$\varepsilon\to0$. The integral over the remaining massive modes
$\delta\sigma$ also yields a constant equal to unity as the numbers of
ordinary variables and Grassmann variables are the same.

What is the contribution of the Efetov--Wegner boundary terms in
Eq.~\eqref{eq:dd:kpointfunction-asymp}? --- We apply Rothstein's
theory~\cite{Roth} to make changes of variables in superspace. Its
main result is that Efetov--Wegner terms can be associated with
certain vector fields, here denoted $\hat Y_o$, see
appendix~\ref{app:Rothstein}.  For the $k$--point correlation
function~\eqref{eq:dd:kpointfunction-asymp}, we change the integration
variables according to $\sigma^{(oo)}\to
T^{(o)}(\delta\sigma_{aa}/\sqrt{p}-\imath
q_0(x^{(o)}+\imath\varepsilon\tilde L^{(o)})){T^{(o)}}^{-1}$. Here,
$\sigma^{(oo)}$ is the $(2m^{(o)}|2m^{(o)})\times(2m^{(o)}|2m^{(o)})$
supermatrix block of $\sigma$ which is at the same position in matrix
space as $T^{(o)}(\delta\sigma_{aa}/\sqrt{p}-\imath
q_0(x^{(o)}+\imath\varepsilon\tilde L^{(o)})){T^{(o)}}^{-1}$. Then
Eq.~\eqref{eq:dd:kpointfunction-asymp} becomes
\begin{align}
 \label{eq:dd:kpointfunction-asymp-b}
 \begin{split}
 \lim_{p\to\infty} R_k(x;\xi)=& \prod_{o=1}^\alpha \biggl[K_{\infty,m_o}\lim_{\varepsilon\rightarrow0}\sum_{L_1^{(o)},\ldots,L_{m_o}^{(o)}=\pm 1}
 \int\exp[-\hat Y_o(T^{(o)},\delta\sigma_{oo})] {\rm d}\mu(T^{(o)})\leb{\delta\sigma_{oo}}\left(\prod_{j=1}^{m_o}\frac{ L_j^{(o)}{\rm Im}\,q_0(x^{(o)})}{8\pi \gamma^2}\right.\\
 &\hspace*{-2.5cm}\times\left.\str T^{(o)}\tilde L^{(o)}{T^{(o)}}^{-1}\left[\begin{array}{cc} e^{m_o}_{jj}&0\\0&- e^{m_o}_{jj}\end{array}\right]\otimes\id_{2}\right)\exp\left(\frac{\imath{\rm Im}\, q_0(x^{(o)})}{2\gamma^2}\Str\tilde\xi^{(o)}T^{(o)}  \tilde L^{(o)} {T^{(o)}}^{-1}-\varepsilon\Str T^{(o)}{T^{(o)}}^\dagger-\frac{l}{4\gamma^2}\str\delta\sigma_{oo}^2\right)\biggl]
 \end{split}
\end{align}
where all Efetov-Wegner boundary terms are taken care of by the vector
fields $\hat Y_o(T^{(o)},\delta\sigma_{oo})$. This is the main motivation to introduce these vector fields. Only with them Eq.~\eqref{eq:dd:kpointfunction-asymp-b} is an equality.  Unfortunately, explicit
expressions for those vector fields are not available in general. Only
for the case of Hermitian supermatrices a successful explicit
identification of all Efetov--Wegner boundary terms was achieved in
Ref.~\cite{guhr1993} at small matrix dimension and in
Ref.~\cite{Kieburg2011} for general supermatrix size. However we are in the lucky position that the vector fields only depend on the change of coordinates but not on the integrand. Thus their explicit expressions are not needed to identify the $k$-point correlation functions of the correlated Wishart ensemble with those of the GOE.

The order of the
action of the operators $\exp[-\hat Y_o(T^{(o)},\delta\sigma_{oo})]$
and the measure ${\rm d}\mu(T^{(o)})d[\delta\sigma_{oo}]$ is important
since ${\rm d}\mu(T^{(o)})$ also incorporates non--trivial
ingredients, see appendix~\ref{app:Rothstein}.  Hence, $\hat
Y_o(T^{(o)},\delta\sigma_{oo})$ does not only act on the integrand but
on this measure, too.

We now can exactly identify the product of
integrals~\eqref{eq:dd:kpointfunction-asymp-b} with the $k$--point
correlation function~\eqref{GOE-k-point}. The vector fields $\hat
Y_o(T^{(o)},\delta\sigma_{oo})$ do fully coincide with those for the
correlated Wishart ensemble because we perform the same change of
integration variables. The integrands are also equal in the large
$p$--limit, apart from the rescaling of the spectral fluctuations
(unfolding), see Eq.~\eqref{unfolding}. We infer the important result
that both correlation functions, including all Efetov-Wegner boundary
terms, are exactly the same. The second equality of
Eq.~\eqref{eq:dd:kpointfunction-asymp} reflects the universality of
the local spectral fluctuations.

A last remark is in order. The factorization of $R_k(x)$ into the
$m^{(o)}$--point correlation functions
$X_{m^{(o)}}(\widehat\xi_1^{(o)},\dots,\widehat\xi_{m^{(o)}}^{(o)})$
does not come as a surprise since we zoom into the spectrum at
different points $x^{(1)},x^{(2)},x^{(3)},\ldots$ Those points are
macroscopically separated such that eigenvalues around $x^{(a)}$ should be
statistically independent from those around another point $x^{(b)}$.
This is so because the other infinitely many eigenvalues in between
cause a screening. The next to leading order in the $1/p$ expansion,
however, must crucially depend on the random matrix model, especially
the confining potential, \textit{e.g.}, see Ref.~\cite{global}.

\section{Outliers and Soft Edges}\label{sec:saddlepoint-edge}

In subsection~\ref{sec:outlier} we investigate the limiting positions
and the fluctuations of possibly existing outliers.  In
subsection~\ref{sec:soft} we derive the exact real Airy kernel
statistics at any soft edge of the bulk.  In
subsection~\ref{sec:large} we trace back the calculation of the
cumulative density function to skew-orthogonal polynomial problem.

\subsection{Outliers}\label{sec:outlier}

An outlier is an eigenvalue that is separated from all other
eigenvalues. It thus suffices to investigate the level
density~\eqref{level-asymp} because the higher correlations involving
outliers are suppressed. We may neglect the outliers in the saddle
point equation~\eqref{ScalarSaddlePoint} for the bulk of the
eigenvalue density because they are $1/p$ corrections, but we have to
study their average position and the width of their distribution. The
peaks in their distribution result from the fact that the saddle point
solution $q_0(x)$ cannot stay on the real line in the vicinity of the
poles $-1/\Lambda_j$. The solution $q_0(x)$ has to leave the real line
when tuning $x$, even though this is only necessary for a very small
interval in $x$.

We consider the outlier $\Lambda_o$, say.  To analyze the behavior of
the saddle point solution in the presence of $\Lambda_o$, we expand
the eigenvalue variable $x=x_0+\delta x/\sqrt{p}$ and the saddle point
$q_0(x)=-1/\Lambda_o+\delta q_0/\sqrt{p}$ in
Eq.~\eqref{ScalarSaddlePoint}. The scaling $1/\sqrt{p}$ for the
deviations, $\delta x$ and $\delta q_0$, will turn out to be the
correct one later on. The variable $\delta x$ probes the level density around the point $x_0$ and, thus, plays the same role as $\zeta$ in Eq.~\eqref{Rk}. The point $x_0$ is the position of the outlier
peak for $p\to\infty$, while its corresponding point  of the saddle point solution is the pole
$q_0(x_0)=-1/\Lambda_o$. To express $x_0$ and $\delta q_0$ as
functions of $\delta x$, we expand the saddle point
equation~\eqref{ScalarSaddlePoint} up to order $1/\sqrt{p}$,
\begin{equation}\label{exp-saddle}
0\approx\left(-x_0+\Lambda_o+\frac{\gamma^2}{p}\sum_{j\neq o}\frac{\Lambda_o\Lambda_j}{\Lambda_o-\Lambda_j}\right)+\frac{1}{\sqrt{p}}\left(-\delta x+\Lambda_o^2\delta q_0-\frac{\gamma^2}{p}\sum_{j\neq o}\frac{\Lambda_o^2\Lambda_j^2}{(\Lambda_o-\Lambda_j)^2}\delta q_0+\frac{\gamma^2}{\delta q_0}\right).
\end{equation}
This expansion is not valid for eigenvalues inside a bulk of
eigenvalues since then the difference $\Lambda_o-\Lambda_j$ might be
less than order one for some $j\neq o$, implying higher order terms in
$p$ in the expansion~\eqref{exp-saddle}. We now see why the above
variations around $x_0$ and $q_0=-1/\Lambda_o$ were chosen of order
$1/\sqrt{p}$ because other dependencies would lead to inconsistent
expansions. Identifying the the terms of order one and $1/\sqrt{p}$
yields two results, namley
\begin{equation}\label{outlier-pos}
 x_0=\left(1+\frac{\gamma^2}{p}\sum_{j\neq o}\frac{\Lambda_j}{\Lambda_o-\Lambda_j}\right)\Lambda_o
\end{equation}
for the limiting position and 
\begin{equation}\label{outlier-saddle}
 \delta q_0(\delta x)=\left(\Lambda_o^2-\frac{\gamma^2}{p}\sum_{j\neq o}\frac{\Lambda_o^2\Lambda_j^2}{(\Lambda_o-\Lambda_j)^2}\right)^{-1}\left(\frac{\delta x}{2}\pm\sqrt{\frac{\delta x^2}{4}-\gamma^2\left(\Lambda_o^2-\frac{\gamma^2}{p}\sum_{j\neq o}\frac{\Lambda_o^2\Lambda_j^2}{(\Lambda_o-\Lambda_j)^2}\right)}\right)
\end{equation}
for the deviation of the saddle point solution from the pole
$q_0=-1/\Lambda_o$.  Interestingly, the position of the outlier is not
directly at $\Lambda_o$ but slightly shifted, \textit{cf.},
Eq.~\eqref{outlier-pos}. Only in the limit $p\gg1$ and
$\Lambda_o\gg\Lambda_j$ for all $\Lambda_j$ in the bulk of the
eigenvalues, we have $x_0=\Lambda_o$.

The fluctuations of the outlier around the position
$x_0$ are of the order
\begin{equation}\label{outlier-deviation}
 \Delta x_0\approx 2\gamma\sqrt{1-\frac{\gamma^2}{p}\sum_{j\neq o}\frac{\Lambda_j^2}{(\Lambda_o-\Lambda_j)^2}}\frac{\Lambda_o}{\sqrt{p}}
\end{equation}
as can be read off from the relation between the level
density~\eqref{level-asymp} and the saddle point $q_0$. The saddle
point only yields a contribution to the spectral density if it has a
non--vanishing imaginary part which, in turn, can only result from the
square root in Eq.~\eqref{outlier-saddle}. This implies a condition on
the empirical eigenvalues for the expansion~\eqref{exp-saddle} to hold,
\begin{equation}\label{condition}
 \frac{\gamma^2}{p}\sum_{j\neq o}\frac{\Lambda_j^2}{(\Lambda_o-\Lambda_j)^2}<1.
\end{equation}
This condition can occasionaly be violated for some time series as we
show in our numerical simulations in section~\ref{numerics}. In such
cases the expansion~\eqref{exp-saddle} fails because the matrix
dimensions are too small. We expect that the
condition~\eqref{condition} is always true for sufficiently large $p$
and $n$ and for an outlier $\Lambda_o$ that is larger than the upper soft
edge of the bulk. This is consistent with the $1/p$ suppression of the
contribution due to other outliers in the sum~\eqref{condition}. If
$p,n$ are too small, the condition~\eqref{condition} fails whenever
$(\Lambda_o-\Lambda_a)^2\leq\Lambda_a^2/n$ with $\Lambda_a$ being
another outlier. The difference $(\Lambda_o-\Lambda_a)$ then has an
order $\sqrt{p}$ behavior, resulting in a higher order polynomial
equation for the saddle points. Another problem arises when the
outlier is too close to a soft edge of a bulk of eigenvalues. Such a
situation can emerge when the noise in the data becomes too strong and
the outliers start to merge with the bulk. Again, the saddle point
equation has to be modified. The worst scenario is when both
situations occur simultaneously.

Another point deserves further discussion. The square root behavior of
the level density, also known from Wigner's semi--circle law, cannot
be interpreted as the limiting distribution of the outlier. A simple
argument from the full random matrix model~\eqref{WishartDist} shows
that, for large $p$ and fixed degeneracy $l$, the distribution for the
outlier around $\Lambda_o$ coincides with the level density of the $l\times l$
Gaussian Orthogonal Ensemble (GOE) centered at $x_0$ and with
fluctuations proportional to $\Delta x_0$. The shape of the outlier level density
encodes the number of eigenvalues, \textit{i.e.}, the degeneracy
$l$, Fig.~\ref{fig:density}. Nonetheless the position
and the widths of the distributions of the outliers are still the
same.  Only in the limit $l\to\infty$ of infinite degeneracy we find
Wigner's semi-circle law again.  The reason for this behavior is the
macroscopic distance of the bulk and the outlier. The distributions of
the individual eigenvalues of the outlier (indeed we have more than
one because of the degeneracy $l$) with those in the bulk will only
have an exponentially small overlap such that one can consider the
eigenvalues in the outlier separately.

Interestingly, the results~\eqref{outlier-pos} and
\eqref{outlier-deviation} seem also applicable to outliers which lie
on a scale different from that of the bulk. The limiting position
$x_0$ as well as the order $\Delta x_0$ of fluctuations scale with
$\Lambda_o$. They are thus likely to become $x_0=\Lambda_o$ and $
\Delta x_0\approx2\gamma\Lambda_o/\sqrt{p}$ for large $p$.

\subsection{Airy Statistics at the Soft Edges}\label{sec:soft}

We restrict ourselves to the soft edges where the asymptotic level
density~\eqref{level-asymp} vanishes as a square root and derive
the Airy kernel. We do not consider higher order multi--critical
points which cannot be excluded \textit{a priori}. For the sake of
simplicity we assume that $x_j=x_0$ for all $j=1,\ldots,k$ coincides
with the position where the level density~\eqref{level-asymp}
vanishes. Then we have only one saddle point
$Q_0=q_0(x_0)\id_{2k|2k}$. We recall that $q_0(x_0)$ starts to become
real at the edge $x_0$ such that it does not have an imaginary part
that is related to the metric $\tilde L$. The scale of the local
fluctuations $\xi$ changes to
$\tilde\kappa=x_0\id_{2k|2k}+\tilde{j}+\xi/(lp)^{2/3}+\imath \tilde
L\varepsilon$, employing the notation of Eq.~\eqref{superZ}. This scale reflects the fact that the level density vanishes like a square root and that one has to expand around the saddle point up to third order.  The
massive modes $\delta\sigma$ around the saddle point solution scales
differently, too, in particular we have the expansion $\sigma=-\imath
q_0(x_0)\id_{2k|2k}+\delta\sigma/(lp)^{1/3}$.  On the scale of the
local fluctuations, we expand the Lagrangian~\eqref{approxLagrangian} up to order $1/p$,
\begin{align}
\label{approxLagrangian.d}
 \begin{split}
&\mathcal L \left(x_0\id_k+\frac{\xi}{(lp)^{2/3}},-\imath q_0\id_{2k|2k}+\frac{\delta\sigma}{(lp)^{1/3}}\right)\\
=&\frac{\imath}{(lp)^{1/3}}\left[ \frac{\gamma^2}{p}\sum_{i=1}^{p} \frac{\Lambda_i}{1+ \Lambda_i q_0}-x_0-\frac{1}{q_0}\right]\Str\delta\sigma+\frac{1}{2(lp)^{2/3}}\left[\frac{\gamma^2}{p}\sum_{i=1}^{p} \frac{\Lambda_i^2}{(1+ \Lambda_i q_0)^2}-\frac{1}{q_0^2}\right]\Str\delta\sigma^2\\
&-\frac{\imath}{3lp}\Str\left[\left(\frac{\gamma^2}{p}\sum_{i=1}^{p} \frac{\Lambda_i^3}{(1+\Lambda_i q_0)^3}-\frac{1}{q_0^3}\right)\delta\sigma^3+3(\tilde\xi+\imath(lp)^{2/3}\varepsilon\tilde L)\delta\sigma\right]+\mathcal{O}\left(\frac{1}{(lp)^{4/3}}\right).
 \end{split}
\end{align}
The term of order $\mathcal{O}(p^0)=\mathcal{O}(1)$ drops out because
the saddle point is proportional to the identity matrix such that the
corresponding supertraces vanish. Moreover the terms of order
$1/(lp)^{1/3}$ and $1/(lp)^{2/3}$ vanish because $q_0(x_0)$ is the
contributing saddle point at the position $x_0$ where the level
density vanishes in a square--root fashion. The first term of
Eq.~\eqref{approxLagrangian.d} is the function $\hat{g}(q(x),x)$ appearing in the
scalar saddle point equation~\eqref{ScalarSaddlePoint}, while the
second term is its derivative $g'(q)$ with respect to the variable
$q$. We underline that the derivative $\partial_q\hat{g}(q,x)=g'(q)$ vanishes at $q_0$, too,
which can be seen as follows. On the one hand, $R_1(x)$ vanishes as
$\sqrt{|x-x_0|}$ such that the Cauchy transform of $R_1(x)$, which is
up to normalization the saddle point solution $q_0$, has a divergent
first derivative at $x=x_0$, \textit{i.e.},
$q'(x)\to1/\sqrt{|x-x_0|}\to\infty$ for $x\to x_0$. On the other hand
the total derivative of the function $\hat{g}(q_0(x),x)$ in the variable $x$
yields $0=d\hat{g}/dx(q_0(x),x)=-1+g'(q_0(x))q'_0(x)$ which indeed has to
vanish because $q_0$ is the saddle point solution. Thus we have
$g'(q_0(x))=1/q'_0(x)$ implying that $g'$ vanishes at $q_0(x_0)$.

The $1/p$ term in Eq.~\eqref{approxLagrangian.d} is the leading term
of the Lagrangian. Thus the $k$-point correlation function at the edge
$x_0$ takes the form
\begin{align}
 \label{eq:dd:kpointfunction-b}
 \begin{split}
 R_k(x_0,(lp)^{1/3}\xi) \leb{(lp)^{1/3}\xi}\overset{p\gg1}{\approx} &K_{n}\lim_{\varepsilon\rightarrow0}  \sum_{L_1,\ldots,L_k=\pm 1}
 \int\leb{\delta\sigma}\prod_{j=1}^k\frac{L_j }{8\pi\imath \gamma^2}\,\str \delta\sigma\left[\begin{array}{cc} e^{k}_{jj}&0\\0&- e^{k}_{jj}\end{array}\right]\otimes\id_{2}\\
 &\times\exp\left(\frac{\imath}{6\gamma^2}\Str\left[\left(\frac{\gamma^2}{p}\sum_{i=1}^{p} \frac{\Lambda_i^3}{(1+\Lambda_i q_0(x_0))^3}-\frac{1}{q_0^3(x_0)}\right)\delta\sigma^3+3(\tilde\xi+\imath(lp)^{2/3}\varepsilon\tilde L)\delta\sigma\right]\right).
 \end{split}
\end{align}
We reiterate that Efetov--Wegner boundary terms and, hence,
non-vanishing vector fields \`a la Rothstein~\cite{Roth} do not
appear. The coordinate transformation is only a constant shift that
cannot cause such contributions.  The terms where we replace
$\delta\sigma$ or its higher powers by the leading order saddle point
solution $\imath q_0(x_0)\id_{2k|2k}$ inside the product of the integrand
in Eq.~\eqref{eq:dd:kpointfunction-b} vanish for the same reason as
in the bulk, see the discussion below
Eq.~\eqref{eq:dd:kpointfunction-asymp}. The integral turns out
invariant under the sub--supergroup ${\rm UOSp}(2|2)$ such that
Cauchy--like Wegner integration
theorems~\cite{Wegner,Con,ConGro,KKG08} apply which reduce the
integral over the supermatrix $\delta\sigma$ to an integral over a
supermatrix of a smaller dimension. Then one of the signs $L_j$ drop out and, thus, the remaining
integrand is independent of the sign of the imaginary increment over
which the sum runs. Precisely this sum yields zero due to the
additional alternating signs $L_j$ in the product.

The limit $\varepsilon\to0$ together with the sign matrix $\tilde L$
and, thus, the original domain of integration of $\sigma$ fixes the
integration contour for the eigenvalues of the boson--boson and
fermion--fermion blocks of $\delta\sigma$. The contour for an
eigenvalue $s_{{\rm B},j}$ in the boson--boson block consists of two
disjoint half--lines and is equal to the union
$-\imath\mathbb{R}_+\cup L_j\mathbb{R}_+$. We emphasize that the
integration over $-\imath\mathbb{R}_+$ results from the negative sign
of $q_0(x_0)$ implying that the saddle point $-\imath q_0(x_0)$ lies
on the positive half--axis. The integrability of $-\imath\mathbb{R}_+$
is ensured by the cubic term, and the integration over
$L_j\mathbb{R}_+$ is absolutely convergent due to the $\epsilon$ term.
When tilting the second half--line to $L_j\mathbb{R}_+\to
(L_j/2+\sqrt{3}\imath/2)\mathbb{R}_+$ we can perform the
$\varepsilon\to0$ limit exactly, because in this more appropriate
integration domain the cubic term dominates on both half--lines. The
integration over an eigenvalue $s_{{\rm F},j}$ in the fermion-fermion
block consists of the two half--lines $e^{\imath
  7\pi/6}\mathbb{R}_+\cup e^{\imath 11\pi/6}\mathbb{R}_+$ independent
of $\tilde L$.

Again we have to unfold the local fluctuations which leads to 
\begin{align}\label{unfolding-edge}
\widehat{\xi}_a^{(o)} =\left(\frac{\gamma^2}{p}\sum_{i=1}^{p} \frac{\Lambda_i^3}{(1+\Lambda_i q_0(x_0))^3}-\frac{1}{q_0^3(x_0)}\right)^{-1/3}\frac{\xi_a}{\gamma^{4/3}}.
\end{align}
To obtain $k$--point correlations, we also have to rescale the
supermatrix $\delta\sigma$ and arrive at
\begin{align}
 \label{eq:dd:kpointfunction-c}
 \begin{split}
 \lim_{p\to\infty}R_k(x_0,(lp)^{1/3}\xi) {\rm d}\left[\xi\right]= &K_{\infty,k}d[\widehat\xi]\sum_{L_1,\ldots,L_k=\pm 1}
 \int\leb{\delta\sigma}\exp\left(\frac{\imath}{6}\Str\left[\delta\sigma^3+3\widehat\xi\delta\sigma\right]\right)\prod_{j=1}^k\frac{L_j }{8\pi\imath}\,\str \delta\sigma\left[\begin{array}{cc} e^{k}_{jj}&0\\0&- e^{k}_{jj}\end{array}\right]\otimes\id_{2}.
 \end{split}
\end{align}
The asymptotic result~\eqref{eq:dd:kpointfunction-c} can be written in
terms of the Airy kernel and is thus equivalent to the result for the
GOE~\cite{TW}. We recall that the GOE is well--known to exhibit Airy
statistics at the soft edges. We choose the local scaling limit of the
corresponding $k$--point correlation functions at the edge $x_0=2$
and find
\begin{align}
\widehat{X}_{k}(\widehat\xi)\leb{\widehat\xi} =&\leb{\widehat\xi}\lim_{\substack{p\to\infty\\ \varepsilon\to0 \\ j\to0}}\sum_{L\in\{\pm 1\}^k}\prod_{i=1}^k \frac{L_i}{4 \imath \pi nl} \partial_{j_i}\frac{\int\leb{H}e^{-nl\,\tr H^2}\Sdet^{-1/2}(H\otimes\id_{2k|2k}-\id_{nl}\otimes(2\id_{2k|2k}+\widehat{\xi}/(nl)^{2/3}+\tilde j+\imath \varepsilon\tilde L))}{\int\leb{H}\exp[-nl\,\tr H^2]}\nonumber\\
=&\leb{\widehat\xi}\lim_{\substack{n\to\infty\\ \varepsilon\to0}}K_{n}\sum_{L\in\{\pm 1\}^k}\int \leb{\sigma}\exp\left(-\frac{nl}{2}\widehat{\mathcal L}(\sigma)\right)\prod_{i=1}^k\frac{L_j }{8}\str\sigma\left[\begin{array}{cc} e^k_{jj}&0\\0&- e^k_{jj}\end{array}\right]\otimes\id_{2}\nonumber\\
=&\lim_{p\to\infty}R_k(x_0,(lp)^{1/3}\xi) {\rm d}\left[(lp)^{1/3}\xi\right]\label{GOE-k-pointedge}
\end{align}
with
\begin{equation}
\widehat{\mathcal L}(\sigma)=\frac{1}{2}\str\sigma^2-\imath\str\left(\imath\varepsilon\tilde{L}+\frac{\widehat{\xi}}{(nl)^{2/3}}\right)\sigma-\left(1-\frac{1}{nl}\right)\str{\rm ln}\,\sigma.
\end{equation}
To arrive at the last equality of Eq.~\eqref{GOE-k-pointedge} we
expanded the supermatrix according to
$\sigma=\imath\id_{2k|2k}+\delta\sigma/(nl)^{2/3}$ and identified the
result with the right hand side of
Eq.~\eqref{eq:dd:kpointfunction-c}. Of course, the edge correlations
of the GOE can be also derived by other methods, \textit{e.g.} orthogonal
polynomials. We conclude that the correlated real Wishart
ensemble~\eqref{WishartDist} shows, at any soft edge
that behave in a square--root fashion, spectral correlations of the
Airy type known from the GOE.

\subsection{Distribution of the Largest Eigenvalue}\label{sec:large}

We consider a particular example to illustrate how useful the
independence of the correlations and densities in the limit of large
matrix dimensions is. In particular, it leads to simpler analytical
results. We study the cumulative density function for the largest
eigenvalue of the correlated Wishart matrix $WW^T$. As we have shown
that the soft edges as well as the outliers are independent of the
generic degeneracy of the empirical correlation matrix $C$ in the
limit of large matrix dimension, we expect that this also hold
approximately for the position and the width of the
largest--eigenvalue distribution. If outliers are not present and the
largest eigenvalue lies at the upper soft edge we find the
Tracy--Widom distribution~\cite{TW}. Later on, we will present
numerical simulations which confirm that, in first approximation, the
distribution for the largest eigenvalue of the bulk of the eigenvalues
indeed shows the expected behavior, see Fig.~\ref{fig:small-large}.

The cumulative density function for the largest eigenvalue of the
correlated real Wishart ensemble with a generic degeneracy $l=2$ in
the empirical correlation matrix $C\otimes\id_2$ is given by
\begin{equation}\label{def-cum-den}
E_{2p,n}(t)=\int \leb{W} P(W|C\otimes\id_2)\Theta(t\id_{2p}-WW^T) \ ,
\end{equation}
where $\Theta$ is the Heaviside function on the symmetric matrices,
\textit{i.e.}, it is unity if the matrix is positive definite and zero
otherwise. The function $E_{2p,n}(t)$ may also be viewed as the gap
probability that none of the eigenvalue is larger than
$t\in\mathbb{R}_+$. Its derivative with respect to $t$ yields the
distribution of the largest eigenvalue.  In
Ref.~\cite{WirtzGuhrKieburgEPL2015} we have shown that such integrals can be mapped to invariant symmetric matrix ensembles. Then the cumulative
density function~\eqref{def-cum-den} can be rewritten as an integral
over a $2 n\times 2n$ real symmetric matrix $H$, namely
\begin{equation}\label{cum-den-alt}
E_{2p,n}(t)=\left(\frac{nt}{2}\right)^{2np}\frac{K'_{n,n}}{\det^{2n} \Lambda}\int \frac{\exp[\tr(\imath H+\id_{2n})]\leb{H}}{\det^{(2n+1)/2}(\imath H+\id_{2n})\prod_{j=1}^p\det(\imath H+(nt\Lambda_j^{-1}/2+1)\id_{2n})}.
\end{equation}
The limit $E_{2p,n}(t\to\infty)=1$ yields the normalization constant
\begin{equation}\label{cum-den-const-a}
\frac{1}{K'_{j,n}}=\int \frac{\exp[\tr(\imath H+\id_{2j})]\leb{H}}{\det^{(2n+1)/2}(\imath H+\id_{2j})}=\prod_{a=0}^{2j-1}\frac{2\pi^{(a+2)/2}}{\Gamma[(2n-a+1)/2]}
\end{equation}
which is a special form of the Ingham-Siegel
integral~\cite{Ingham,Siegel}.

Without the degeneracies, the square roots of the determinants
$\det(\imath H+(nt\Lambda_j^{-1}/2+1)\id_{2n})$ in the
integrand~\eqref{cum-den-alt}, {\it cf.}
Ref.~\cite{WirtzGuhrKieburgEPL2015}, would be most cumbersome. Luckily,
the double degeneracy of each empirical eigenvalues combines two
square roots and yields a determinant to power one. This is a
considerable advantage as compared to the non--degenerated
case. Hence, we can algebraically reformulate the integrand such that
the integral drastically simplifies. To this end, we diagonalize the
matrix $H=OEO^T$ with $O\in{\rm SO}(2n)$ and
$E=\diag(E_1,\ldots,E_{2n})\in\mathbb{R}^{2n}$,
\begin{equation}\label{cum-den-1}
E_{2p,n}(t)=\left(\frac{nt}{2}\right)^{2np}\frac{\tilde{K}_{n,n}}{(2n)!\det^{2n} \Lambda}\int \frac{\exp[\tr(\imath E+\id_{2n})]|\Delta_{2n}(E)|\leb{E}}{\det^{(2n+1)/2}(\imath E+\id_{2n})\prod_{j=1}^p\det(\imath E+(nt\Lambda_j^{-1}/2+1)\id_{2n})}
\end{equation}
where the intergration over the orthogonal group~\cite{MehtasBook} leads
to the new normalization constant
\begin{equation}\label{cum-den-const-b}
\frac{1}{\tilde{K}_{j,n}}=\frac{1}{K'_{j,n}}\prod_{a=0}^{2j-1}\frac{2\Gamma[(a+3)/2]}{\pi^{(a+1)/2}}=\prod_{a=1}^{2j}\frac{4\sqrt{\pi}\Gamma[(a+2)/2]}{\Gamma[(2n-a+2)/2]}=\prod_{a=1}^{j}\frac{2^{2n-4a+5}\pi(2a)!}{(2n-2a+1)!}
\end{equation}
Algebraic rearrangement~\cite{KG12} and the usage of
skew--orthogonal polynomials~\cite{skew-2,skew-1} uncovers the
Pfaffian structure of the integral~\eqref{cum-den-1},
\begin{equation}\label{cum-den-2}
E_{2p,n}(t)=\frac{\tilde{K}_{n,n}}{\tilde{K}_{n-p/2,n}}\frac{{\rm Pf}\left[\mathcal{K}_{n}(nt\Lambda_a^{-1}/2,nt\Lambda_b^{-1}/2)\right]_{a,b=1,\ldots,p}}{\det(2\Lambda/(nt))^{2n}\Delta_p(nt\Lambda^{-1}/2)}
\end{equation}
with the kernel
\begin{align}
\mathcal{K}_{n}(x_1,x_2)=&\frac{1}{\imath}\biggl(\int\frac{{\rm d}E_1{\rm d}E_2\,{\rm sign}(E_1-E_2)}{(\imath E_1+x_1+1)(\imath E_2+x_2+1)}\frac{\exp(\imath E_1+\imath E_2+2)}{(\imath E_1+1)^{(2n+1)/2}(\imath E_2+1)^{(2n+1)/2}}\nonumber\\
&-\sum_{l=0}^{p/2-1}\frac{\widehat{q}_{2l}(x_1)\widehat{q}_{2l+1}(x_2)-\widehat{q}_{2l}(x_2)\widehat{q}_{2l+1}(x_1)}{h_l}\biggl).\label{cum-den-kern-1}
\end{align}
This result is only true for $p$ even. For $p$ odd we may augment the
empirical eigenvalues $\Lambda$ with a dummy eigenvalue
$\Lambda_{p+1}$ such that we effectively extend $p\to p+1$ and
eventually take the limit $\Lambda_{p+1}\to\infty$. We refrain from
showing the details and stick to the case of $p$ even in the sequel.

The functions $\widehat{q}_l(x)$ in Eq.~\eqref{cum-den-kern-1} are the
Cauchy transforms
\begin{equation}\label{Cauch-pol}
\widehat{q}_{l}(x)=\int\frac{{\rm d}E_1{\rm d}E_2\,{\rm sign}(E_1-E_2)q_l(E_1)}{\imath E_2+x+1}\frac{\exp(\imath E_1+\imath E_2+2)}{(\imath E_1+1)^{(2n+1)/2}(\imath E_2+1)^{(2n+1)/2}},
\end{equation}
of the skew--orthogonal polynomials $q_l(E)$ (in monic normalization)
according to
 \begin{equation}\label{skew-ortho}
   \int{\rm d}E_1{\rm d}E_2\,{\rm sign}(E_1-E_2)q_{2a}(E_1)q_{2b+1}(E_2)\frac{\exp(\imath E_1+\imath E_2+2)}{(\imath E_1+1)^{(2n+1)/2}(\imath E_2+1)^{(2n+1)/2}}=h_a\delta_{ab},
 \end{equation}
 with $a,b\in\mathbb{N}_0$. All other bilinear relations between the
 polynomials vanish. The constants $h_a$ follow from the normalization
 constant~\eqref{cum-den-const-b},
 \begin{equation}\label{norm-h}
 \frac{1}{\tilde{K}_{j,n}}=\prod_{l=0}^{j-1}h_l \qquad \longleftrightarrow \qquad
 h_j=\frac{\tilde{K}_{j,n}}{\tilde{K}_{j+1,n}}=\frac{2^{2n-4j+1}\pi(2j-2)!}{(2n-2j-1)!}.
 \end{equation}
 The Cauchy transform $\widehat{q}_l(x)$ is readily derived as a
 Heine--type--of formula~\cite{AKP10}
 \begin{equation}\label{Heine-1}
 \widehat{q}_{2l}(x)=\frac{h_l K'_{l+1,n}}{\imath^{2l+1}(2l+2)!}\int \frac{\exp[\tr(\imath H+\id_{2l+2})]\leb{H}}{\det^{(2n+1)/2}(\imath H+\id_{2l+2})\det(\imath H+(x+1)\id_{2l+2})}
 \end{equation}
 and
 \begin{equation}\label{Heine-2}
 \widehat{q}_{2l+1}(x)=-\frac{h_l K'_{l+1,n}}{\imath^{2l}(2l+2)!}\int \frac{(x+\imath\tr H+c_l)\exp[\tr(\imath H+\id_{2l+2})]\leb{H}}{\det^{(2n+1)/2}(\imath H+\id_{2l+2})\det(\imath H+(x+1)\id_{2l+2})}
 \end{equation}
 with an arbitrary constant $c_l$ which cannot be fixed with the
 skew--orthogonality relation but can be used as a gauge parameter. The matrix $H$ is a $(2l+2)\times(2l+2)$ real symmetric matrix.

 The integral~\eqref{Heine-1} is very similar to the gap probability
 $E_{p=1,n}(t=1)$ at $t=1$, \textit{cf.}, Eq.~\eqref{cum-den-alt},
 with the empirical correlation matrix $C^{-1}\to 2x/n$, in particular
 we have
 \begin{equation}\label{Heine-1a}
 \widehat{q}_{2l}(x)=\frac{(-1)^l h_l}{\imath \pi^{2n}}\ x^{2(n-l)-2}\int \leb{W} \exp[-x\tr WW^T]{\det}^{2n-2l-2}(\id_2-WW^T)\Theta(\id_2-WW^T)
 \end{equation}
 with $W$ a $2\times 2n$ real matrix. The Cauchy transform $\widehat{q}_{2l+1}(x)$ of the odd polynomials
 can also be expressed in terms of such an integral, as it can be
 traced back to a derivative of $\widehat{q}_{2l}(x)$,
 \begin{equation}\label{Heine-2a}
 \widehat{q}_{2l+1}(x)=-\imath\left(x+c_l-2\imath(l+1)+2(l+1)(n-l)+x\frac{\partial}{\partial x}\right)\widehat{q}_{2l}(x).
 \end{equation}
 Setting $c_l=2\imath(l+1)-2(l+1)(n-l)$ we have
 \begin{align}\label{Heine-2b}
   \widehat{q}_{2l+1}(x)=&-\imath x\left(1+\frac{\partial}{\partial x}\right)\widehat{q}_{2l}(x)\\
   =&\frac{(-1)^{l+1} h_l}{ \pi^{2n}}\ x^{2(n-l)-1}\int \leb{W}
   \left(1-\tr WW^T\right)\exp[-x\tr
   WW^T]{\det}^{2n-2l-2}(\id_2-WW^T)\Theta(\id_2-WW^T),\nonumber
 \end{align}
 where $W$ is a real $2\times 2n$ matrix. We point out that the
 imaginary unit in Eq.~\eqref{Heine-1a} cancels with the one in the
 kernel~\eqref{cum-den-kern-1} such that the result is indeed as
 required. The integral~\eqref{Heine-1a} can be evaluated in closed
 form by diagonalizing the $2\times 2$ Wishart correlation matrix
 $WW^T$ and integrating over the corresponding two eigenvalues. This
 yields the finite sum
 \begin{equation}
 \widehat{q}_{2l}(x)=d_l\sum_{b=0}^{2(n-l-1)}\binom{2(n-l-1)}{b}\frac{\,_2F_1(3/2-n,1+b;2+b;-1)}{(2^{n+1/2}-2)(1+b)}\frac{\,_1F_1(2(b-2n+2l+2);4l-6n+4;2x)}{x^{4n-2l-2}},\label{Heine-1b}
 \end{equation}
 with the constant $d_l=(-1)^lh_l(6n-4l-5)!/[\imath 2^{4(n-l-1)}(2n-2)!]$. The functions $\,_1F_1$ and $\,_2F_1$ are the confluent and Gauss'
 hypergeometric functions, respectively. The functions
 $\widehat{q}_{2l+1}(x)$ can be evaluated via
 relation~\eqref{Heine-2b}, we omit the details.
 
 Altogether, we derived the rather simple and fairly explicit
 results~\eqref{cum-den-2}, \eqref{cum-den-kern-1}, \eqref{norm-h},
 \eqref{Heine-2b} and \eqref{Heine-1b} for the cumulative distribution
 of the largest eigenvalue of $WW^T$ in the presence of degeneracies
 in $C$ ($l=2$), \textit{cf.}
 Ref.~\cite{WirtzGuhrKieburgEPL2015}. Without degeneracies,
 non--trivial analytical problems arise due to the square roots of
 determinants. Applying now our observation that the spectral
 statistics become the same for large time series, our above results
 asymptotically solve the corresponding problem without degeneracies.
 Hence, we developed a general method to obtain asymptotic results
 for other quantities of the correlated
 real Wishart ensemble by  artificially introducing double
 degeneracies in the empirical correlation matrix $C$.
\begin{figure}[t!]
 \centering
 \includegraphics[width=0.8\textwidth]{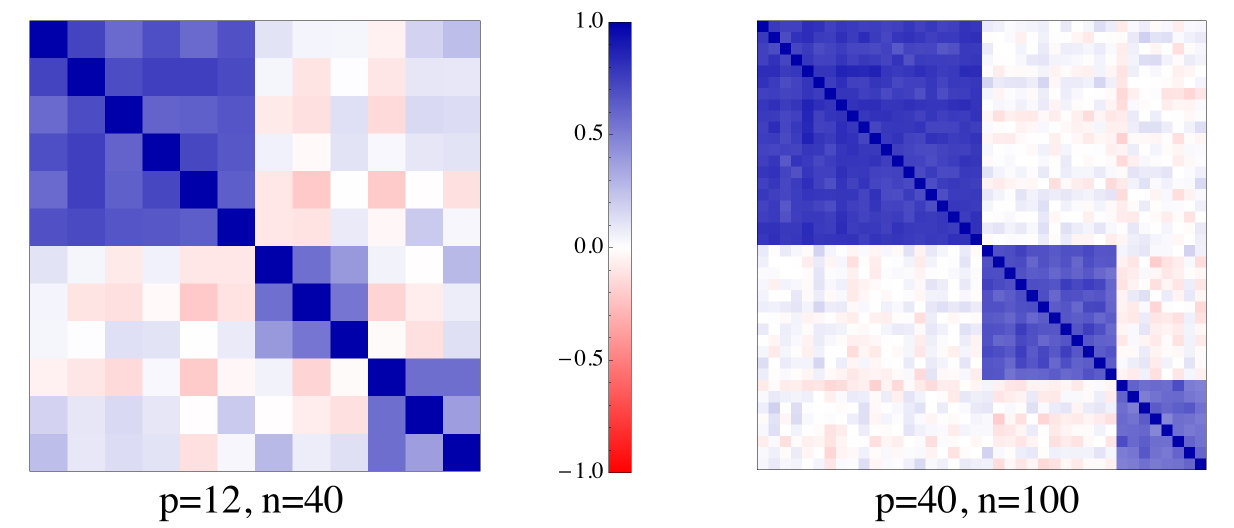}
 \caption{The two empirical correlation matrices of a $12\times40$ time
   series (left plot) and a $40\times100$ time series (right
   plot) which were employed for the Monte Carlo simulations. The strength of the correlation is color coded as shown in the legend.}
\label{fig:EmpCorI}
\end{figure}

\section{Numerical Simulations}\label{numerics}

\begin{figure}[t!]
 \centering\includegraphics[width=1\textwidth]{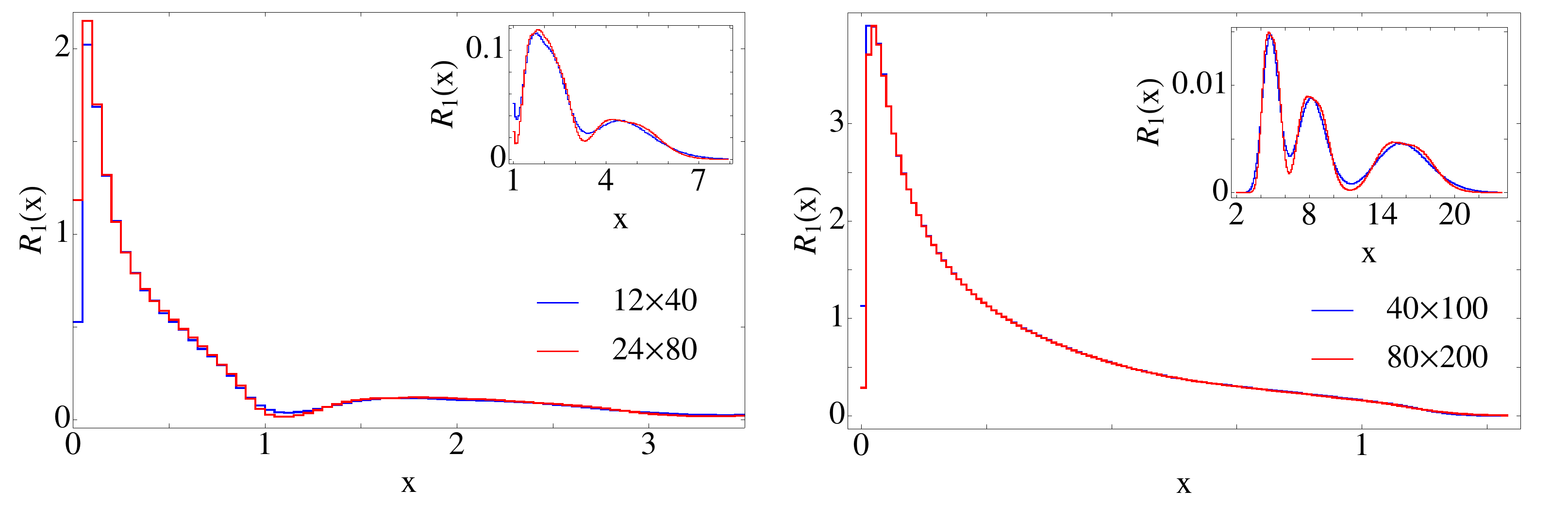}
 \caption{Level densities as histograms for the real Wishart ensembles
   with the two empirical correlation matrices shown in
   Fig.~\ref{fig:EmpCorI}. Blue lines correspond to the
   non--degenerated and red lines to the doubly degenerate empirical
   correlation matrices. The level densities around the outliers
   are shown on a magnified scale in the insets.}
 \label{fig:density}
\end{figure}

For illustrating purpose and to show the robustness of our
approximations and predictions, we carry out two Monte Carlo
simulations of the correlated real Wishart
ensemble~\eqref{WishartDist}. We use a one--factor model, see
\textit{e.g.} Ref.~\cite{ross,noh00}, to construct two sets of time series
$T_{12\times40}$ ($p=12$ and $n=40$) and $T_{40\times100}$ ($p=40$ and
$n=100$). Each set $T=T_0+s_{\rm noise} T_1$ consist of a signal $T_0$
featuring three perfectly correlated sectors and a fully uncorrelated
white--noise offset $T_1$ such that $\langle \{T_1\}_{ab}\rangle=0$
and $\langle
\{T_1\}_{ab}\{T_1\}_{a'b'}\rangle=\delta_{aa'}\delta_{bb'}$. The
strength of the noise is tuned by the parameter $s_{\rm noise}$.  In
the simulations we choose $s_{\rm noise}=3$ for $T_{12\times40}$ and
as $s_{\rm noise}=4$ for $T_{40\times100}$. From these sets of times
series we derive the corresponding empirical correlation matrices
$C_{12\times40}$ and $C_{40\times100}$. They are shown in
Fig.~\ref{fig:EmpCorI}.  The three strongly correlated sectors show up
as deep blue blocks on the diagonal although the white noise is of the
same order as the signal. The sizes of these blocks, $(6,3,3)$ for
$T_{12\times40}$ and $(20,12,8)$ for $T_{40\times100}$, mainly determine the
positions of the three largest eigenvalues (outliers) of the corresponding
empirical correlation matrices, $\Lambda_{12\times40}^{\rm
  (out)}\approx\diag(4.44,2.17,2.03)$ for $T_{12\times40}$ and
$\Lambda_{40\times100}^{\rm
  (out)}\approx\diag(15.61,8.39,5.08)$ for  $T_{40\times100}$. However, we see strong shifts in Fig.~\ref{fig:density} (left) for the smaller time series $T_{12\times40}$ because of the relatively strong noise and the relatively small matrix dimensions.

We numerically simulate the real Wishart ensemble for each of these
two so constructed empirical correlation matrices $C_{12\times40}$ and
$C_{40\times100}$ and their doubly degenerate counter parts
$C_{12\times40}\otimes\id_2$ and
$C_{40\times100}\otimes\id_2$. Altogether we simulate four
ensembles. The ensembles consist of $10^6$ matrices for each empirical
correlation matrix. These large ensemble sizes lead to high
statistical significance.  In Fig.~\ref{fig:density}, we present the
macroscopic level densities including outliers, the statistical errors
amount to a few percents at most.  The level densities employing the
degenerate and non--degenerate empirical correlation matrices show
perfect agreement in the bulk of the empirical eigenvalues. Not
surprisingly, the agreement is better for larger dimension
$p$. Nevertheless, even for low matrix dimensions $p$ and $n$, the
deviations in the bulk are small.  At the edges and for the outliers
the deviations become visible beyond the statistical error. They
result from the statistical fluctuations of the individual eigenvalues
around their average positions due to the level repulsion caused by the overlapping tails of the individual eigenvalue distributions which are still present at finite matrix dimension. In the bulk the eigenvalues
are more abundant, implying that their respective positions are
sharper. In contrast, the eigenvalues near the soft edges explore the
region outside the limiting support, while they strongly accumulate at
the hard edge as the cross--over to the negative real line is
forbidden. This behavior is suppressed by a generic degeneracy in the
empirical correlation matrix. Although the empirical correlation
matrix might be degenerate, the corresponding Wishart correlation
matrix $WW^T$ is not. Hence there are for the doubly degenerate matrix
twice as many eigenvalues in $WW^T$ as in the non-degenerate
case. This implies that the degenerate case is closer to the
asymptotic result~\eqref{level-asymp} derived by the saddle point
solution. In particular, the support becomes more restrictive. The
same discussion also applies to the outliers whose overlaps with the
other eigenvalues are more suppressed when the empirical correlation
matrix is doubly degenerate. We notice that the level densities around
the outliers only reaches values of up to two orders smaller than in
the bulk.

\begin{figure}[t!]
 \centering
 \includegraphics[width=1\textwidth]{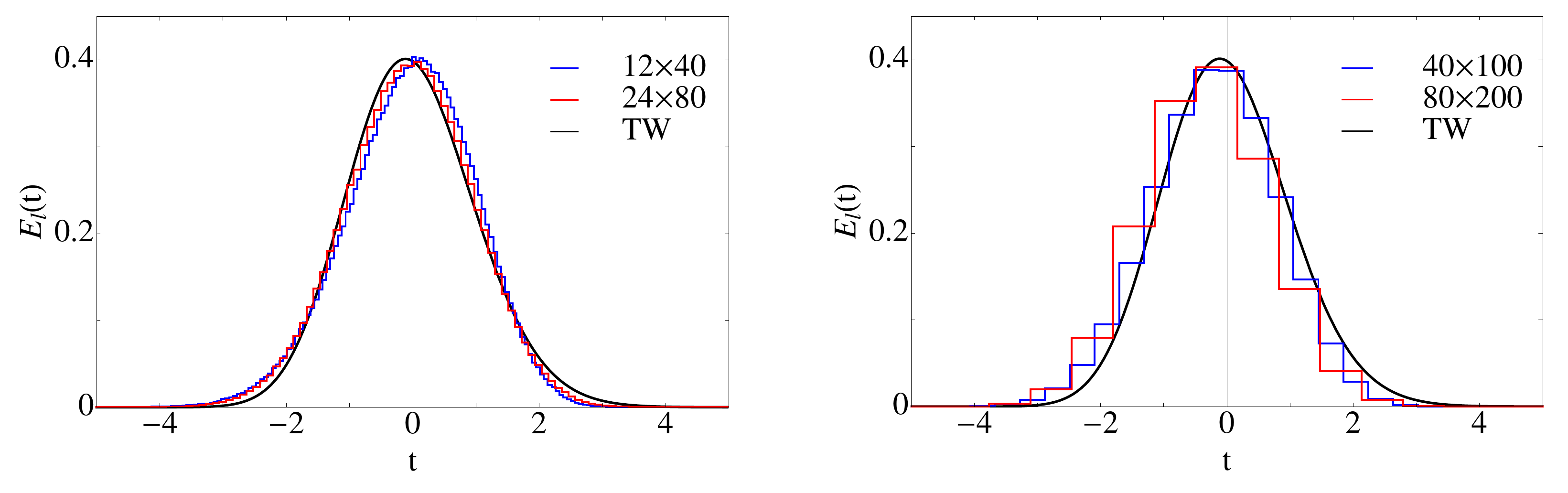}
 \centering
 \includegraphics[width=1\textwidth]{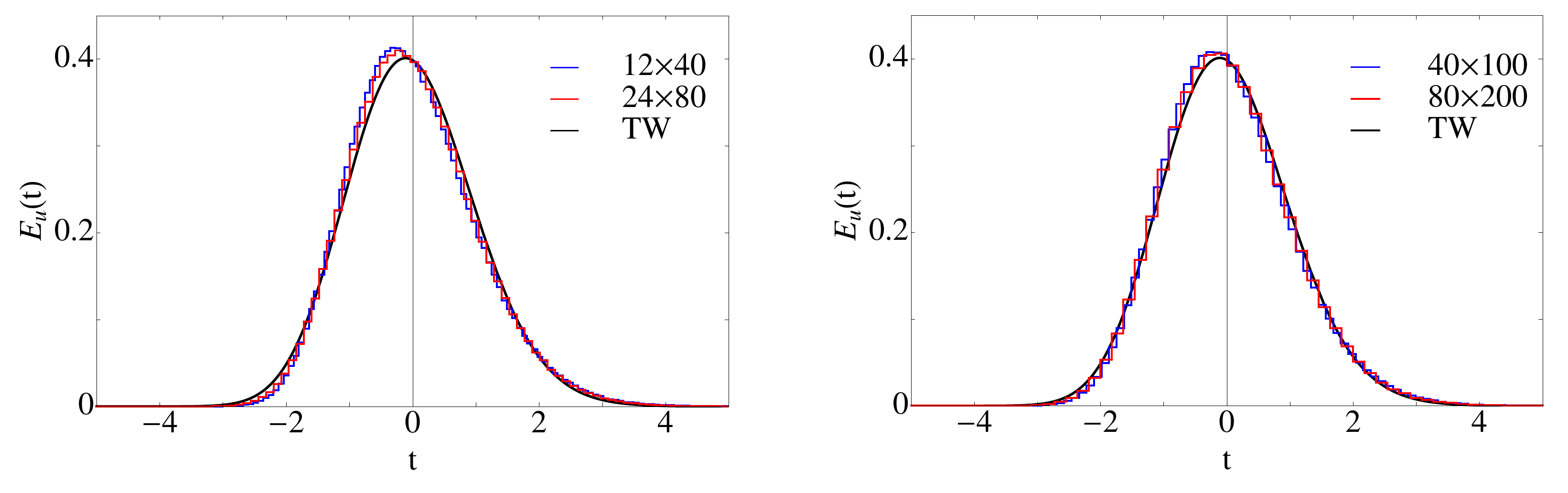}
 \caption{The distributions of the smallest ($E_{\rm l}(t)$, top row)
   and of the largest ($E_{\rm u}(t)$, bottom row) eigenvalues of the
   bulk normalized to zero mean and variance one. We consider again
   the same ensemble of Fig.~\ref{fig:density} with the $12\times40$
   correlation matrix (left column) and the $40\times100$ correlation
   matrix (right column) of Fig.~\ref{fig:EmpCorI}. The histograms for
   the non-degenerate (blue) and the doubly degenerate (red) empirical
   correlation matrix are also compared to approximations~\eqref{TW}
   for the Tracy-Widom distribution (black smooth curve, TW) for real
   matrices. The agreement with the limiting Tracy-Widom distribution
   is good regarding the small matrix dimensions and even the leading order in the deviations
   from this distribution seem to be independent of the degeneracy.}
 \label{fig:small-large}
\end{figure}

Although the level density of the bulk exhibits the strongest
differences at its edges, the spectral statistics on the local scale
converges surprisingly well for the Wishart ensembles with and without
the degeneracies in the empirical correlation matrices.  This is seen
in Fig.~\ref{fig:small-large} which displays the
distribution of the largest and smallest eigenvalue at the edges of
the bulk. For the comparison, the numerical results are unfolded such that the distributions have zero mean and unit
variance. Moreover, the distributions of the smallest eigenvalue are
mirrored at the origin to compare the numerical
results with the Tracy--Widom distribution~\cite{TW} for real matrices
which should be the limiting distribution for large matrix dimensions
$p$ and $n$. The Tracy--Widom distribution indicates that the Airy
statistics holds in this regime. We employed the approximation 
\begin{equation}\label{TW}
 E_{\rm TW}(t)\approx 6.68\times 10^{-76} (t+8.93)^{78.66}\exp(-8.93 t) \ , \qquad t>-8.93 \ ,
\end{equation}
of the Tracy--Widom distribution~\cite{TW-approx}. Again, this
distribution was normalized to zero mean and unit variance for simpler comparison. This means we shifted the distribution given by Chiani~\cite{TW-approx} by the mean and rescaled it by its standard deviation.  The
agreement with the Tracy--Widom distribution is quite good despite the
small matrix dimensions $p=12,40$ in our numerical simulations.  The
more important result, however, is the good agreement of the two
distributions for the degenerate and for the non--degenerate empirical
correlation matrices.  We also mention that even the leading order
deviations of the numerical simulations from the limiting
distribution~\eqref{TW} seem to be approximately independent of the
degree of the degeneracy $l$ in the empirical correlation matrix.

The influence of the degeneracy in the empirical correlation matrix is
strongest for the level density around the outliers, see the insets in
Fig.~\ref{fig:density}. The reason was already discussed at the end of
subsection~\ref{sec:outlier}. The number of eigenvalues associated to
each outlier is equal to the degeneracy, namely $l$. Hence, the shape
of the distribution for each outlier strongly depends on $l$. However
the mean value and the standard deviations of the distributions around
the outliers should not change much with the degeneracy. To leading
order we expect an independence which indeed is confirmed by the
numerical simulations.

\begin{figure}[t!]
 \centering\includegraphics[width=1\textwidth]{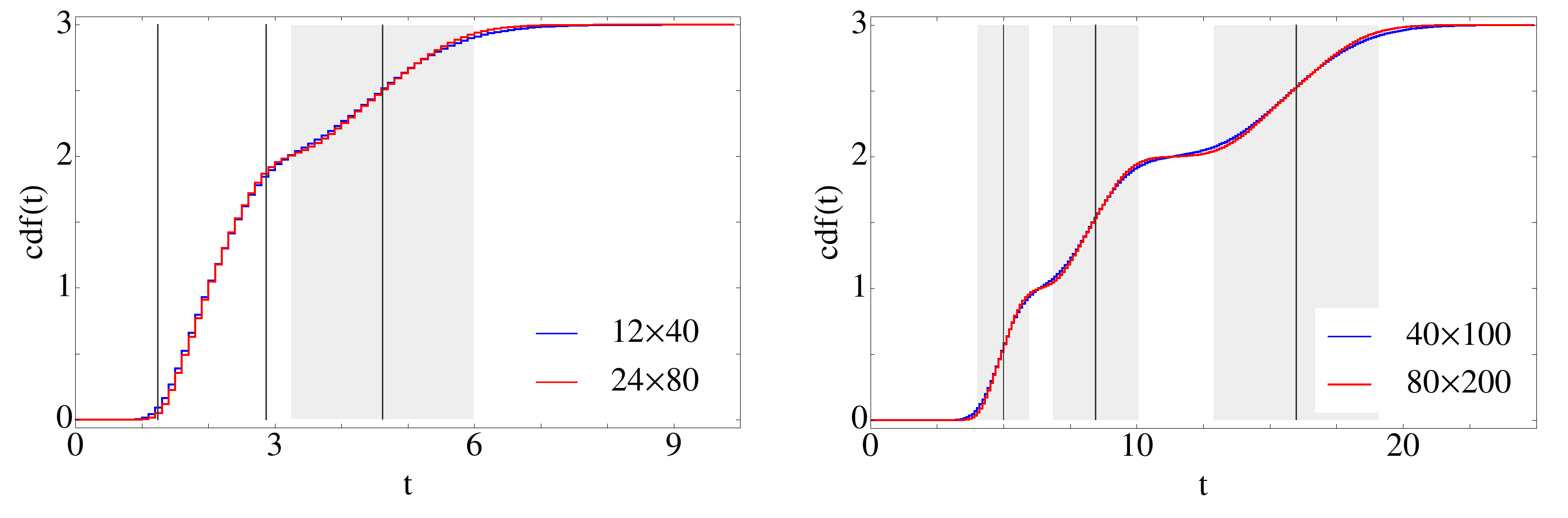}
 \caption{Cumulative density functions ${\rm cdf}(t)$ around the three
   outliers for the real Wishart ensembles with the empirical
   correlation matrices shown in Fig.~\ref{fig:EmpCorI}, for the time series
   $T_{12\times40}$ (left) and $T_{40\times100}$
   (right). Blue and red histograms for the non--degenerate and degenerate
   empirical correlation matrices, respectively. Black vertical lines
   indicate the predicted positions~\eqref{outlier-pos} of the
   outliers and the grey shaded areas are the predicted
   fluctuations~\eqref{outlier-deviation}. The predicted fluctuations for the smallest outliers for the time series $T_{12\times40}$ have imaginary values such that they have no grey shaded areas.}
 \label{fig:outlier}
\end{figure}

In Fig.~\ref{fig:outlier} the cumulative
distribution function ${\rm cdf}(t)$ is depicted. Being independent of
the bin size, it provides a better measure than the distribution
itself.  The agreement with the analytical prediction of the
positions~\eqref{outlier-pos} and the
fluctuations~\eqref{outlier-deviation} for the three outliers is
almost perfect for the set of the larger time series $T_{40\times100}$
and thus seen to be independent of the degree of degeneracy. This also
holds for the largest outlier in the case for the set of the smaller
time series $T_{12\times40}$, while the two smaller outliers do not
follow at all the analytical predictions. For the fluctuations of
these two eigenvalues we find imaginary values with
Eq.~\eqref{outlier-deviation}, indicating that the approximation
discussed in subsection~\ref{sec:outlier} fails. The reason becomes
clear when looking at the inset of the left plot in
Fig.~\ref{fig:density}. The two outliers overlap too much and even
start to merge with the bulk. Hence, one has to modify the
approximation presented in subsection~\ref{sec:outlier}, as discussed
below Eq.~\eqref{condition}. Nonetheless the difference in the
cumulative distributions of the outliers for the smaller and larger
time series differ only marginally for the non--degenerate and
degenerate case. This underlines our claim that even the outliers are
in leading order unaffected by the (artificial) degeneracy.

\section{Conclusions}\label{conclusion}

Our study has produced three main results. The first one is that the
spectral statistics of a real Wishart ensemble with a given empirical
correlation matrix are independent of an artificially introduced
degeneracy of the empirical eigenvalues.  We derived this under
moderate assumptions on the empirical correlation matrix and for an
arbitrary degree of degeneracy. It holds for the local as well as for
the macroscopic bulk statistics. Surprisingly, even the positions and
the width of the fluctuations of possible outliers are independent of
the degeneracy. The differences between the non--degenerate and the
degenerate cases are the strongest close to the edges of the bulk and
in the shape of the distribution around the outliers
statistically significant differences between the non--degenerate and
the degenerate cases emerge. We explained this behaviour
theoretically and confirmed it with Monte--Carlo simulations.

The second main result is that the bulk and soft--edge statistics on
the local scale of the mean level spacing follows the one of the
Gaussian Orthogonal Ensemble (GOE). As we used the
supersymmetry technique, we had to handle Efetov--Wegner boundary
terms. We solved this problem employing Rothstein's theory and thereby
exactly identified the statistics in the correlated real Wishart
ensemble and in the GOE. Performing numerical simulations, we were
able to compare the distribution of the largest and the smallest
eigenvalue of the bulk with the Tracy--Widom distribution. The
agreement is remarkably good even for small matrix dimensions.

Our third main result is a proposition, strongly corroborated by our
analytical findings. As the degeneracies in the empirical correlation
matrices do not influence the spectral statistics in a relevant
fashion, we suggest to study the doubly degenerate case of an
empirical correlation matrix instead of the non--degenerate one when
one wishes to derive asymptotic analytical results for observables
such as the distributions of individual eigenvalues and the level
density. Due to the absence of square roots of determinants in the
integrands, the doubly degenerate case is by far easier to treat than
the non--degenerate one.  As an example we employed results of
Ref.~\cite{WirtzGuhrKieburgEPL2015} for the cumulative density function
of the largest eigenvalue and derived an expression in terms of a
Pfaffian in which all integrals are evaluated in closed form. We
expect that other spectral observable can be asymptotiacally computed
as well with this new method. Of course, for finite number and length
of the time series this approach only yields an approximation, but our
numerical simulations indicate that these approximations are quite
good even for relatively small matrix dimensions.

\begin{acknowledgments}
  We acknowledge support from the Deutsche Forschungsgemeinschaft,
  Sonderforschungsbereich TR12 (T.W. and T.G.),
  Sonderforschungsbereich 701 (M.K.) and the Alexander von Humboldt
  Foundation (M.K.). We considerably benefitted from the inspiring
  atmosphere at the conference ``Random Matrix Theory: Foundations and
  Applications'' in Cracow, Poland, where we started this project in
  July 2014.
\end{acknowledgments}

\appendix

\section{Rothstein's Theory for Boundary Terms in Superanalysis}
\label{app:Rothstein}

We consider an arbitrary diffeomorphism mapping one coordinate system
$(y,\eta)$ of a superspace to another one $(x,\theta)$. Here, we
employ the notation of Rothstein~\cite{Roth}, implying that
$(x,\theta)$ and $(y,\eta)$ should not be confused with variables we
use in the body of the paper.  The transformation of an integral over
an arbitrary superfunction $f$ is not purely given by the Berezinian
(Jacobian) but also incorporates corrections, henceforth abbreviated
``${\rm b.t.}$", the so called Efetov-Wegner boundary terms,
\begin{equation}
\int f(y,\eta) d[y,\eta]=\int f(y(x,\eta),\eta(x,\theta)) \sdet\left(\frac{\partial(y,\eta)}{\partial(x,\theta)}\right)d[x,\theta]+{\rm b.t.}.
\end{equation}
One can control these boundary terms by splitting the diffeomorphism
into two steps. First we map the coordinate system to the numerical
part $y_0$ of $y$ and to the first order part (in the Grassmann
variables $\theta$) $\eta_1$ of $\eta$. We denote the intermediate
coordinates $(x',\theta')$ such that
\begin{equation}
\int f(y,\eta) d[y,\eta]=\int f(y_0(x'),\eta_1(x',\theta')) \sdet\left(\frac{\partial(y_0,\eta_1)}{\partial(x',\theta')}\right)d[x',\theta'].
\end{equation}
This transformation is free of Efetov--Wegner boundary terms because
$y_0$ does not contain any Grassmann variables $\theta$. In the next
step we can generate the remaining diffeomorphism by a vector field
$\hat{Y}(x,\theta)$ via
$(y(x,\eta),\eta(x,\theta))=(y_0(x'(x,\eta)),\eta_1(x'(x,\theta),\theta'(x,\theta)))=\exp[\hat{Y}(x,\theta)](y_0(x),\eta_1(x,\theta))$
which yields the full transformation formula
\begin{equation}
\int f(y,\eta) d[y,\eta]=\int \exp[-\hat{Y}(x,\theta)]f(y_0(x),\eta_1(x,\theta)) \sdet\left(\frac{\partial(y_0,\eta_1)}{\partial(x',\theta')}(x,\theta)\right)d[x,\theta].
\end{equation}
The correctness of this procedure was proven in \cite[Chapter 3]{Roth}.

Two properties are known of the vector field $\hat{Y}$. First, it is a
nilpotent vector field and a sum of even orders in the Grassmann
variables. Thus the operator $\exp[-\hat{Y}]$ is a finite sum of
powers of $\hat{Y}$ with the maximal power equal to half of the number
of Grassmann variables. In our problem it would be $2k^2$ and hence
independent of the dimensions $p$ and $n$.  The second property of the
vector field is that it only depends on the coordinate transformation
and not on the integrand. We make use of this property in our
calculation when identifying the $k$--point correlation function of
the correlated Wishart ensemble with the sine kernel for the GOE.

\end{document}